\documentclass[pre]{revtex4}
\usepackage{epsfig}
\begin{document}
\def\be{\begin{equation}}
\def\ee{\end{equation}}
\def\bfi{\begin{figure}}
\def\efi{\end{figure}}
\def\bea{\begin{eqnarray}}
\def\eea{\end{eqnarray}}
\newcommand{\ket}[1]{\vert#1\rangle}
 \newcommand{\bra}[1]{\langle#1\vert}
\def\id{\mathbb{I}}
\title{Correction to scaling in the response 
function of the $2d$ kinetic Ising model}

\author{Federico Corberi$^\ddag$, Eugenio Lippiello$^\dag$ 
and Marco Zannetti$^\S$}
\affiliation {Istituto Nazionale di Fisica della Materia, Unit\`a
di Salerno and Dipartimento di Fisica ``E.Caianiello'', 
Universit\`a di Salerno,
84081 Baronissi (Salerno), Italy}

\begin{abstract}

The aging part $R_{ag}(t,s)$ of the impulsive response function  of the two dimensional 
ferromagnetic Ising model, quenched below the critical point, is studied numerically
employing a new algorithm without the imposition of the external field.
We find that the simple scaling form $R_{ag}(t,s)=s^{-(1+a)}f(t/s)$, which is usually
believed to hold in the aging regime, is not obeyed. 
We analyse the data assuming the existence of a correction to scaling.
We find $a=0.273 \pm 0.006$, in agreement with previous 
numerical results obtained from the zero field cooled magnetization.
We investigate in detail also the scaling function $f(t/s)$
and we compare the results with the predictions of analytical theories.
We make an ansatz for the correction to scaling, deriving an analytical
expression for $R_{ag}(t,s)$. This gives a satisfactory qualitative agreement
with the numerical data for $R_{ag}(t,s)$ and for the integrated response functions.
With the analytical model we explore the overall behavior, extrapolating beyond the time regime accessible 
with the simulations. We explain why the data for the zero field cooled
susceptibility are not too sensitive to the existence of the correction to scaling in $R_{ag}(t,s)$,
making this quantity the most convenient for the study of the asymptotic scaling
properties.

\end{abstract}
\maketitle

\ddag corberi@sa.infn.it  
\dag lippiello@sa.infn.it 
\S zannetti@na.infn.it

\vspace {0.5cm}


\section{Introduction}\label{intro}

A powerful tool in the study of slowly relaxing systems is the extension of the 
fluctuation dissipation theorem (FDT) to off-equilibrium conditions.
The relation between the response and the autocorrelation function has been shown 
to encode basic properties of the dynamics~\cite{Parisi99} 
and of the equilibrium state~\cite{Franz98}. 
However, while the autocorrelation
function $C(t,s)$ has been studied since a long time and, in some systems,
is well understood, the knowledge of the response function $R(t,s)$
remains comparatively poor. 

In the prototypical case, considered here, of non disordered coarsening systems 
quenched below the critical temperature, 
general properties of $C(t,s)$ and $R(t,s)$ may be inferred
from the structure of the configurations.
It is well known, in fact, that in the late stage of phase ordering 
the interior of the growing domains is equilibrated, while their boundary
is out of equilibrium. Accordingly,
a distinction between bulk and interface fluctuations can be made. 
For the autocorrelation function this leads to the 
splitting~\cite{Franzvirasoro,ninfty,Bouchaud97,Culeshouches}
\be
C(t,s) = C_{st}(t-s) + C_{ag}(t,s) 
\label{Ag.1}
\ee
where $C_{st}(t-s)$ is the contribution from the interior of domains, 
behaving as the equilibrium autocorrelation function $C_{eq}(t-s)$ of the ordered 
state at the final temperature $T$ of the quench, while $C_{ag}(t,s)$ is the 
off-equilibrium aging contribution, coming from the interfaces.
Similarly, for the response function one has
\be
R(t,s) = R_{st}(t-s) + R_{ag}(t,s)
\label{Ag.2}
\ee
and the stationary contributions $R_{st}(t-s)$ and $C_{st}(t-s)$ are related by the FDT
\be
TR_{st}(t-s)=\frac {\partial C_{st}(t-s)}{\partial s}.
\label{fdt}
\ee
For $s$ large, $C_{ag}(t,s)$ obeys the simple aging scaling form
\be
C_{ag}(t,s)=s^{-b}g(x)
\ee
with $b=0$ and $x=t/s$~\cite{Bray94,Culeshouches,JSTAT}. 
Taking $s=0$ and $t$ large, $C_{ag}(t,0)$
decays algebraically as $C_{ag}(t,0)\sim t^{-\lambda /z}$, where $\lambda $ 
is the Fisher-Huse exponent~\cite{FisherHuse} and $z$ is the
dynamic exponent regulating the growth of the characteristic size $L(t)\sim t^{1/z}$ 
of domains. Assuming that the same behavior holds also for $s>0$~\cite{cop} gives
$g(x)\sim x^{-\lambda /z}$, for $x\gg 1$. 
These properties of $C(t,s)$, as mentioned above, are well understood and well documented both
by analytical and numerical results.

By analogy, the response function is also expected to obey, for large $s$, a simple aging scaling form 
\be
R_{ag}(t,s)=s^{-(1+a)}f(x)
\label{scalr}
\ee
with
\be
f(x)\sim x^{-\lambda_R /z}
\label{vagiu}
\ee
for $x\gg 1$.
Evidence that this is the case comes from the few models tractable by analytical 
methods~\cite{ninfty}~\cite{Lippiello2000,Godreche2000,epj,Berthier,Mazenko04}. 
In the large majority of cases, where information can be extracted only from numerical simulations,
investigating $R(t,s)$ is considerably more difficult than $C(t,s)$. 
The reason is that  $R(t,s)$ is much more noisy than $C(t,s)$. A possible way around this
difficulty is to study the less noisy
integrated response functions (IRF), such as the zero field cooled magnetization (ZFC) 
\be
\chi(t,t_w) = \int_{t_w}^t ds R(t,s)
\label{0.1}
\ee
or the thermoremanent magnetization (TRM)
\be
\rho(t,t_w) = \int_0^{t_w} ds R(t,s).
\label{0.2}
\ee 
However, if resorting to the IRF does indeed cut down the noise, 
it has turned out that recovering $R(t,s)$ from the IRF is a delicate task which involves more than one 
subtlety (see Ref.~\cite{resteso} and Section~\ref{sectexponent} of this paper).
As a result, so far, in the literature no consensus has been reached neither on the exponent $a$ nor
on the form of the scaling function $f(x)$. Therefore, it seems that the issue of the properties of $R(t,s)$
can be settled only by going to its direct measurement. Recently, this has become feasible
after the introduction of new algorithms~\cite{Chatelain03,Ricci,Eugenio}
for the computation of the response function without applying the external
perturbation. This speeds the simulation to such an extent that 
the direct measurement of $R(t,s)$ has become accessible~\cite{Chatelain03,Ricci,Eugenio,Abriet}.

In this paper we present results for $R(t,s)$
in the two dimensional Ising model obtained with our algorithm~\cite{Eugenio}.
The outcome is quite rich and interesting. We find the unexpected
result that in the $d=2$ Ising model
the simple scaling form~(\ref{scalr}) is not enough to represent the data, 
but a large correction term is needed, up to the
longest times we have reached in the simulations. The quality of the data allows to detect and analyze this correction and,
by taking it properly into account, to make statements on the exponent $a$ 
and the scaling function $f(x)$ appearing in Eq.~(\ref{scalr}).

The paper is organized as follows: In Section~\ref{sectexponent} we summarise the problem of the
exponent $a$. In Section~\ref{algo}
we recall the basic features of the
measurement of the response function without applying the external
perturbation and the method we use to isolate
$R_{ag}(t,s)$ in Eq.~(\ref{Ag.2}). In Section~\ref{results}
we present and discuss the numerical results for $R_{ag}(t,s)$. In particular,
we show the existence of a large preasymptotic scaling correction, we
extract the value of the exponent $a$ and we analyze the scaling function
$f(x)$. 
In Section~\ref{irf}, we investigate the implications of the structure of $R_{ag}(t,s)$
on the properties of the IRF and we compare our predictions with
the numerical simulations of these quantities. This gives further insight into
the problem of the retrieval of $R(t,s)$ from the IRF.
Finally, in Section~\ref{concl} we draw the conclusions.

\section{The exponent $a$}\label{sectexponent}

Before entering the presentation of data and results, we recall briefly what 
is the problem with the exponent $a$ in Eq.~(\ref{scalr}).

\begin{itemize} 

\item The straightforward substitution of Eq.~(\ref{scalr}) into Eqs.~(\ref{0.1}) and~(\ref{0.2}) yields
the scaling forms of the aging parts of the IRF
\be
\chi_{ag}(t,t_w)=t_w^{-a_{\chi}}F_{\chi}(y)
\label{2.1}
\ee
and
\be
\rho_{ag}(t,t_w)=t_w^{-a_{\rho}}F_{\rho}(y)
\label{2.2}
\ee
with
\be
a_{\chi}=a_{\rho}=a.
\label{2.3}
\ee
and $y=t/t_w$.

\item an intuitively appealing argument, originally introduced by Barrat~\cite{Barrat98}, predicts
that $\chi_{ag}(t,t_w)$ ought to be proportional to the 
density of defects at the time $t_w$, which goes as $L(t_w)^{-n}$, with 
$n =1$ and $n =2$ for scalar and vectorial order parameter, respectively. Using  Eq.~(\ref{2.3}),
this leads to the dimensionality independent result
\be
a_{\chi}=a_{\rho}=a=n/z
\label{2.4}
\ee
since, we recall, the exponent $z$ in the quenches below $T_C$ does
not depend on $d$. For systems without conservation of the order parameter, as considered in this paper, $z=2$ ~\cite{Bray94}.

\item contrary to the previous statement, {\it all} available analytical results, which include the exact solutions of the large $N$
model~\cite{ninfty} and of the $d=1$
Ising model~\cite{Lippiello2000,Godreche2000}, as well as 
approximate calculations~\cite{epj,Berthier,Mazenko04}, show that $a$ depends linearly on dimensionality according to
\be
a=\frac {n}{z}\frac {d-d_L}{d_U-d_L}
\label{aphen}
\ee
where $d_L$ is the lower critical dimension of the statics and $d_U > d_L$ is a dimensionality whose significance becomes clear
as soon as ZFC is considered. In fact,
from these analytical calculations it comes out that the relation between $a_{\chi}$ and $a$ is not as simple 
as in Eq.~(\ref{2.3}). Rather, it must be replaced by
\be
    a_{\chi} = \left \{ \begin{array}{ll}
        a  \qquad $as in Eq.~(\ref{aphen}) for$ \qquad d < d_U  \\
	n/z   \qquad $with logarithmic corrections for$ \qquad d = d_U\\ 	
        n/z   \qquad $for$ \qquad d > d_U. 
        \end{array}
        \right .
        \label{2.5}
\ee 
This behavior of $a_{\chi}$ is due to the existence of an irrelevant variable in  $\chi_{ag}(t,t_w)$, which becomes dangerous
for $d \geq d_U$~\cite{ninfty,resteso}. In the large $N$ model the above formula holds with ($n=2,d_L=2,d_U=4$),
while in the approximate calculations with scalar order parameter of Refs.~\cite{epj,Berthier,Mazenko04} ($n=1,d_L=1,d_U=2$).

\item The problem of the exponent $a$, then, is whether Eq.~(\ref{aphen}) is a peculiarity of just those cases
where analytical results are available, the generic behavior being that of Eq.~(\ref{2.4}), {\it or}, viceversa, it is
Eq.~(\ref{aphen}) that captures the generic behavior, revealing thereafter that the argument leading to Eq.~(\ref{2.4}) does to miss
some important feature in the mechanism of the response.

\end{itemize}

\noindent In order to answer the question one has to investigate as many systems as possible, with the aim of putting together
the generic picture. This can be done only by numerical methods. We have carried out such a program 
performing simulations and measuring $a_{\chi}$ in several systems with conserved and 
non conserved dynamics, both with scalar and vectorial order parameter, at different dimensionalities~\cite{Castellano2004}.
The large body of results that we have obtained does indeed indicate, quite convincingly in our opinion, that 
Eq.~(\ref{2.5}) is of general validity, with $d_U=3$~\cite{nota*} and $d_U=4$ for systems with scalar and vector order 
parameter, respectively. 

This conclusion is challanged, mainly on the basis of the measurement of $a_{\rho}$ from TRM in the Ising model with 
$d=2$ and $d=3$~\cite{Henkel2001}, which seems to agree with  Eq.~(\ref{2.4}). In other words, the investigations of ZFC and TRM 
seem to produce
different results. We have explained in detail elsewhere~\cite{resteso} that the difference is only apparent and is due to the fact
that while the data for ZFC are asymptotic, those for TRM are not. To us, the interesting question remaining open is not anymore
what is the value of $a$, but what is the physics behind Eq.~(\ref{aphen}). 
However, as of yet this is not a shared conclusion~\cite{HPPA,CLZ05,reply}. 
So, what emerges from this brief account of the problem is that working with the IRF is a hairy business, because
it has opened the somewhat intricate problem of why ZFC and TRM seem to give different results.
Therefore, in order to make progress, a fresh starting is needed. In this respect,
the direct numerical study of $R(t,s)$, with the new algorithms, seems well suited to the task. 
We shall concentrate on the $d=2$ Ising model, where discrimination between
Eqs.~(\ref{2.4}) and~(\ref{aphen}) ought to be easier, since the difference in the predicted values of $a$ is quite large
\be
    a = \left \{ \begin{array}{ll}
        1/2  \qquad $from Eq.~(\ref{2.4})$   \\ 	
        1/4  \qquad $from Eq.~(\ref{aphen})$. 
        \end{array}
        \right .
        \label{2.6}
\ee

\section{The algorithm}\label{algo}

\subsection{Measurement of the response without applying a perturbation}

The basic idea, in the new methods~\cite{Chatelain03,Ricci,Eugenio} for the measurement of the response function without applying the
external perturbation, is to relate $R(t,s)$ to
some correlation function of the unperturbed dynamics,
much in the same way as in the equilibrium FDT. In this paper we will use our
method introduced in~\cite{Eugenio}, since we have checked
that it is numerically more efficient. Let us briefly
describe it, referring to~\cite{Eugenio} for details.
  
We consider a spin system with Hamiltonian ${\cal H}[\sigma]$,
where $[\sigma]$ is a generic configuration of the spin variables
$\sigma_i=\pm 1$.
The response function is defined by 
\be
R(t,s)=\lim_{\Delta s \to 0} \frac{1}{\Delta s}
\left . \frac{\partial \langle \sigma_i(t)\rangle }{\partial h_i} \right
\vert_{h=0} 
\label{4}
\ee
where $h_i$ is a magnetic field acting on the $i$-th site during 
the time interval $[s,s+\Delta s]$, and the right hand side does not depend on
$i$ due to space translational invariance. 
Computing $\langle \sigma _i(t)\rangle$ by means of the master equation and 
inserting the result into Eq.~(\ref{4}), one arrives at
\be
T R(t,s)=\frac{1}{2}\lim _{\Delta s \to 0}\left [
\frac {C(t,s+\Delta s)-C(t,s)}{\Delta s}-\langle \sigma _i(t-\Delta s)B_i(s)\rangle \right ]
\label{treb}
\ee
where $B_i$ enters the evolution of the magnetization according to
\be
\frac {d\langle \sigma _i(t)\rangle}{dt}=\langle B_i(t)\rangle .
\ee  
This result holds in complete generality for generic Hamiltonian
and transition rates, with or without conservation of the order parameter. In the single spin flip dynamics 
\be 
B_i(t)=2\sigma _i(t)w([\sigma ]\to [\sigma _i])
\ee
where $w([\sigma ]\to [\sigma _i])$ is the transition rate between
two configurations differing for the value of the spin on the site $i$.
In the simulations 
time is discretized by single
updates which, measuring time in montecarlo steps, occur on the microscopic time scale $\epsilon =1/N$.
The best numerical approximation to the limit $\Delta s \to 0$
in Eq.~(\ref{treb}) is obtained by taking
$\Delta s =\epsilon $, which gives
\be
T R(t,s)=\frac{1}{2}\left [
\frac {C(t,s+\epsilon)-C(t,s)}{\epsilon}-\langle \sigma _i(t-\epsilon)B_i(s)\rangle \right ].
\label{trebb}
\ee
In addition to $R(t,s)$, in the following we will also be interested in the general IRF defined 
by~\cite{resteso}
\be
\mu(t,[\tilde{t},t_w])=\int _{t_w}^{\tilde{t}} R(t,s)ds
\label{irff}
\ee
which corresponds to the application of the perturbation between
the times $t_w$ and $\tilde{t} \leq t$. For this quantity, 
Eq.~(\ref{trebb}) is replaced by
\be
T\mu (t,[\tilde{t},t_w])=\frac{1}{2}\left [C(t,\tilde{t})
-C(t,t_w)\right ]-\frac{\epsilon}{2}
\sum _{s=t_w}^{\tilde{t}} \langle \sigma _i(t-\epsilon)B_i(s)\rangle ,
\label{trec}
\ee 
where the sum $\sum _{s=t_w}^{\tilde{t}} $
is over the discrete times in the interval $[t_w,\tilde{t}]$.
  
Despite the  advantages of the method, the computation of the
impulsive response $R(t,s)$ remains a very
demanding numerical task.
In order to improve the signal to noise ratio, therefore, we have
found convenient to consider, instead of $R(t,s)$, the quantity
$\mu (t,[s+\delta,s])$ with $\delta/s  \ll 1$.
Assuming for $R(t,s)$ the scaling form (\ref{scalr}), to first order in
$\delta/s$ one has
\be
\mu (t,[s+\delta,s])=  R(t,s)\left [\delta -\frac{\delta^2}{s}f_1(x)+...\right ]
\ee
with $f_1(x)=(1/2)[(1+a)+ x \frac{d\ln f(x)}{dx}]$.
For $\delta=1$, $\mu (t,[s+\delta,s])$ differs from $R(t,s)$  
by a correction of order $1/s$, which we have checked to be always
negligible in our simulations.
Therefore, in the following, whenever results
for $R(t,s)$ will be presented, it is understood that the data are obtained
with this procedure, if no other specification is made.

In the following we shall use the above formulas for the $d=2$ Ising model with 
nearest neighbors interaction and evolving with Glauber dynamics.

\subsection{The aging contribution}

In this Section we discuss an auxiliary dynamics, referred to as no-bulk-flip (NBF), which is 
used~\cite{resteso,epj,Castellano2004,first} 
to isolate the aging part $R_{ag}(t,s)$ of the response function
appearing in Eq.~(\ref{Ag.2}).

We introduce a classification of
the degrees of freedom by making the distinction between bulk and
interface spins. 
A spin $\sigma_i$ is regarded as 
belonging to the  bulk of a domain if it is aligned with all its nearest neighbors. 
In the NBF algorithm bulk spins cannot flip. 
With this rule the interior
of domains orders rapidly and all what is left is interface dynamics.
Since in coarsening kinetics the aging contribution
comes exclusively from the boundary of domains, measuring quantities 
in a simulation with the NBF rule yields the aging behavior.

\begin{figure}
    \centering
   \rotatebox{0}{\resizebox{1\textwidth}{!}{\includegraphics{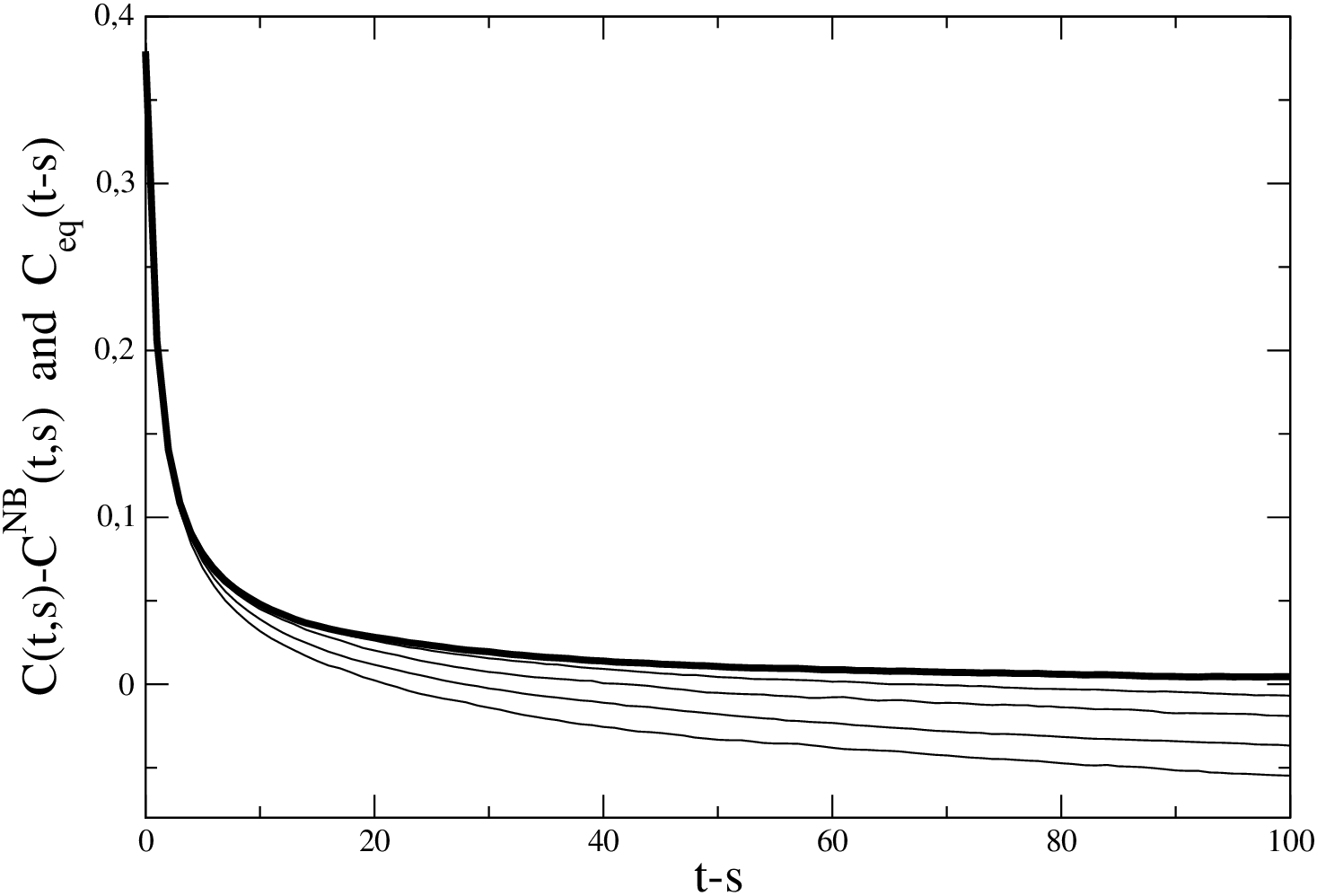}}}
    \caption{Plot of the difference $C(t,s)-C^{NB}(t,s)$ for
            $s=6\cdot 10^2, 2\cdot 10^3, 10^4, 5\cdot 10^4$  (from bottom to top) showing the convergence toward
            $C_{eq}(t-s)$ ( bold curve) as $s$ increases.}
 
\label{fig1}
\end{figure}

In order to illustrate this idea, let us consider the
autocorrelation function. 
Denoting with $\sigma_i^{NB}(t)$ the value of the
spin evolving with NBF dynamics, the quantity
\be
C^{NB}(t,s) = M^2 \langle \sigma_i^{NB}(t) \sigma_i^{NB}(s) \rangle 
\label{Ag.3}
\ee
where $M$ is the equilibrium spontaneous magnetization at the temperature
$T$, is expected to coincide with $C_{ag}(t,s)$.
The $M^2$ factor in front of the definition~(\ref{Ag.3}) is needed
recalling~\cite{ninfty,Bouchaud97,Culeshouches} that $C_{ag}(t,s)$ falls from $M^2$ (the Edwards-Anderson order parameter) 
to zero. 
According to the discussion of Sec.~\ref{intro},
$C_{st}(t-s)$ coincides with the equilibrium correlation $C_{eq}(t-s)$, which
decays from $1-M^2$ to zero.
In Fig.\ref{fig1} we have made the comparison between the
stationary parts of $C(t,s)$ computed in two different ways.
In the first case we measure directly $C_{eq}(t-s)$  
in the equilibrium state prepared at the temperature $T/J=2.2$ ($M^2\simeq 0.616$). We recall that the critical temperature
is given by $T_C/J=2.26918$, where $J$ is the nearest neighbors coupling constant. 
The second prescription is of  
calculating the autocorrelation function $C^{NB}(t,s)$ in the quench
from an initial disordered state,
corresponding to an infinite temperature, to the same final temperature $T$ with the NBF rule, 
and then subtracting it from $C(t,s)$, computed with the full dynamics.
If the NBF algorithm correctly isolates the aging part of the dynamics
what is left is the stationary term, 
$C(t,s)-C^{NB}(t,s)=C_{eq}(t-s)$.
However, one does not expect this to be fulfilled at all times.
The reason is that a sharp separation between two independent components, 
bulk and interface, applies only when $L(t)$ is 
sufficiently large~\cite{ninfty,MVZ}.
Actually, the larger the domains are, the larger is the average distance from a
bulk spin to the nearest interface, so that they are more effectively
decoupled.  
The data of Fig.~\ref{fig1} show a convergence of $C(t,s)-C^{NB}(t,s)$
toward $C_{eq}(t-s)$ as $s$ increases, in agreement with the previous discussion. 
This implies that, for large $s$, the NBF algorithm does to 
isolate the aging part. 

One arrives at the same conclusion by considering the
response function. 
The stationary part $R_{eq}(t-s)$ decays to zero
on a characteristic microscopic time. Then, by using $s$ and $x$ as independent
variables (which we will do from now on) and denoting  with $R(x,s)$ the response function
in terms of these variables, in the limit of large $s$
the stationary part gives a contribution only for $x\simeq 1$.
Therefore, for sufficiently large $s$ and $x>1$, one has
$R(x,s)\simeq R_{ag}(x,s)$. If the NBF algorithm isolates
the aging contribution, the response $R^{NB}(x,s)$
obtained with the NBF rule and $R(x,s)$, measured using the
usual dynamics, should coincide for large $s$.
In order to check this, we have computed 
$R(x,s)$ and $R^{NB}(x,s)$ for
a system of $1000^2$ spins, quenched from the initial disordered state,
corresponding to an infinite temperature, to the final temperature
$T/J=1.5$ ($M^2 \simeq 0.9732$). We have considered several values of $s$ 
ranging from $s=100$ to $s=1600$ and
times $t>s$ up to $5000$. More precisely, we have used (and we shall use in the following) values of $s$ generated from
$s_n = 100 +$Int($1.5^n)$ with $n$ ranging from $1$ to $18$.
The range of times considered
belongs to the scaling regime of the system.
The results in Fig.~\ref{fig2} show that 
$R(x,s)$ and $R^{NB}(x,s)$ coincide within 
the statistical errors. Small differences can be detected only for 
the smallest values of $s$. This is expected since, as 
anticipated, the agreement improves with increasing $s$.

In conclusion, the analysis of both the autocorrelation and the
response function demonstrates the reliability of the NBF algorithm.
Moreover, this algorithm has the advantage of speeding up considerably
the simulations, since only the fraction of spins on
the interfaces must be updated. All the results presented in the following have been obtained
with this technique.

\section{Scaling of $R_{ag}(x,s)$} \label{results}

The first task is to check the scaling properties of $R_{ag}(x,s)$. If Eq.~(\ref{scalr}) was obeyed
the curves for $s^{1+a}R_{ag}(x,s)$, obtained for different values of $s$ and a suitable $a$,
should collapse on a single master curve. However, a rough inspection of Fig.~\ref{fig2} already shows that
this is not the case, since the various curves cannot be superimposed by a 
simple vertical translation. 
We make this more precise in Fig.~\ref{fig3}, where one can see clearly
that there is no collapse for neither one of the two values of $a$ proposed in Eq.~(\ref{2.6}).
For $a=1/4$ the collapse is rather good for small $x$, but
gets worst with increasing $x$. For $a=1/2$ the collapse is bad everywhere,
except for $x\simeq 4$. 

\subsection{The effective exponent $a_{eff}^R(x,s)$ and the value of $a$} \label{effect}

In order to make these considerations quantitative we introduce the effective exponent defined by
\be
1+a_{eff}^R(x,s) = - \left. {\partial \ln R_{ag}(x,s) \over \partial \ln s} \right |_{x}.
\label{Sc.01}
\ee 
Numerically, for a chosen value of $x$, $a_{eff}^R(x,s)$ is given by the local slope of the
plot of $\ln R_{ag}(x,s)$ against $\ln s$ in a selected interval $I_s$ around
$s$. This interval must be chosen small with
respect to the range of $s$ over which $a_{eff}^R(x,s)$ varies appreciably.
However, the smaller the interval $I_s$ the more noisy gets $a_{eff}^R(x,s)$.
On the basis of the data available from the simulations, the best compromise between the needs of
having a local quantity and of lowering the noise has been reached by taking $I_s$
spanning over four consecutive values of $s$~\cite{nota4}. The result is displayed in Fig.~\ref{fig4a},
where symbols with error bars represent the numerical values of $a_{eff}^R(x,s)$, obtained
for the three different $I_s$ with the four values of $s$ indicated in the legend. 
The continuous curves have been obtained from a fitting procedure, which will be discussed in 
Sec.~\ref{fit}. If Eq.~(\ref{scalr}) did hold, we should have found a flat effective exponent, i.e.
$a_{eff}^R(x,s)= a$ independent of $x$ and $s$. Instead, the data for $a_{eff}^R(x,s)$ show the following features:

\begin{enumerate}

\item {\it Fixed $s$}

For fixed $s$, there is a strong dependence on $x$, revealing that Eq.~(\ref{scalr}) is not obeyed.
Furthermore, the curves display a discontinous behavior at $x=1$. The equal times value at $x=1$, ranging
from  $a_{eff}^R(1,s)=-0.55$ to  $a_{eff}^R(1,s)=-0.53$, depending on the $I_s$ considered, 
is separated by a jump from 
the smoothly increasing curve starting around $a_{eff}^R(1^+,s)= 0.30$. By $x=1^+$ we denote the smallest value of $x>1$
used in the simulations. For larger values of $x$, $a_{eff}^R(x,s)$ keeps growing 
continuously. The longest set of $x$ data, corresponding to $I_s$ with the smallest values of $s$, shows the
possible saturation to an asymptotic finite value $a_{eff}^R(\infty,s) > a_{eff}^R(1^+,s)$.
This behavior of $a_{eff}^R$ explains immediately why the attempts to collapse the data with a fixed
value of $a$, as in Fig.~\ref{fig3}, do fail except in a restricted range of $x$.

\item {\it Fixed $x$}  

The size of the error bars makes it difficult to detect an $s$ dependence for fixed $x$, except in the region of short
time difference $x \leq 2$, where with the first two $I_s$ the decrease of $a_{eff}^R(x,s)$ upon increasing $s$
exceeds the error (Fig.~\ref{fig4b}). For larger values of $x$, error bars overlap and no statement can be made. 

\end{enumerate}

\noindent Therefore, Eq.~(\ref{scalr}) is not obeyed and we make the assumption that 
the deviation is due to a correction to scaling of the form
\be
R_{ag}(x,s)=R_1(x,s)+R_2(x,s)=s^{-(1+a)}f(x)+s^{-(1+c)}h(x)
\label{2scalr}
\ee
where necessarily $c>a$, if the new term has to be subdominant. Having made this assumption, from Eq.~(\ref{Sc.01}) follows
\be
1+a_{eff}^R(x,s)=(1+a) \left [ \frac{ 1 + \frac {(1+c)}{(1+a)} s^{-(c-a)}\kappa(x)}
{1+ s^{-(c-a)}\kappa(x)} \right ]
\label{aeffanall}
\ee
where
\be
\kappa(x)= h(x)/f(x).
\label{kappa}
\ee
Let us now pause to explore the consequences of
this formula.   From Eq.~(\ref{aeffanall}) follows that $a_{eff}^R(x,s) \rightarrow a$ from above as 
$s \rightarrow \infty$, for any fixed $x$. As pointed out previously, for those values of $x$ where 
the size of errors is small enough,
the decrease of $a_{eff}^R(x,s)$ with increasing $s$ is confirmed by
the simulations (Fig.~\ref{fig4b}).   Therefore, the smallest measured value of $a_{eff}^R(x,s)$ overestimates $a$.
Excluding the value at $x=1$, for the reasons which will be explained in Sec.~\ref{erre1}, we have
\be
a \leq a_{eff}^R(1^+) = 0.32 \pm 0.01 
\label{exponent}
\ee
where for $a_{eff}^R(1^+)$ we have taken the value at $x=1^+$, corresponding to the intermediate $I_s$ set.
Comparing with Eq.~(\ref{2.6}), we find that this result is compatible with Eq.~(\ref{aphen}) and rules out Eq.~(\ref{2.4}).
This conclusion has been reached with the sole hypothesis that a correction to scaling term needs to
be taken into account. We have made no assumptions neither on the form of $h(x)$ nor on the value of $c$, except for
the obvious requirement $c > a$. In a short while we shall refine considerably the above estimate of $a$, obtaining a
value much more close to the $1/4$ predicted by Eq.~(\ref{aphen}).

\subsection{Analysis of $R_1(x,s)$ and $R_2(x,s)$} \label{erre1}

Let us go further with the analysis of Eq.~(\ref{aeffanall}), identifying the properties of the scaling functions $f(x)$ and $h(x)$, 
which must be obeyed in order
to reproduce the observed features of $a_{eff}^R(x,s)$. Notice that $R_{ag}(x,s)$, being a response function, 
must vanish for large $x$. This requires
that also $f(x)$ and $h(x)$ must vanish for large $x$. Therefore, the saturation to the finite asymptotic value  
$a_{eff}^R(\infty,s) > a_{eff}^R(1^+,s)$,
which is suggested by Fig.~\ref{fig4a},
can occur only if $\kappa(x)$ diverges for large $x$, that is, if $f(x)$ decays faster than $h(x)$.
This has interesting consequences. First of all we get
\be
a_{eff}^R(\infty,s)=c
\label{c}
\ee
independent of $s$. Then, the correction to scaling contribution $R_2(x,s)$ is subdominant for fixed $x$ and large $s$, but 
becomes dominant for fixed $s$ and large $x$.
Which of the two contributions $R_1(x,s)$ and $R_2(x,s)$ is dominant and which is subdominant, depends on the
choice of which variable is kept fixed and which is let to grow. More precisely, the condition
\be
{R_1(x,s) \over R_2(x,s)} = {s^{(c-a)} \over \kappa(x)} = 1
\label{ratio}
\ee
defines the crossover curve $\overline x(s)$ which divides the $(s,x)$ plane (Fig.\ref{fig5}) in the two regions $\Gamma_1$
and  $\Gamma_2$, below and above $\overline x(s)$, where either $R_1(x,s)$ or $R_2(x,s)$ is dominant.
Therefore, if in the simulations one could reach values of $s$ and $t$ so large to have 
$\overline x (s)\gg 1$, together with a range of $x$ extending well beyond $\overline x (s)$, 
one should observe a neat crossover from $a_{eff}^R(x,s)=a$, in a wide interval $1<x\ll\overline x(s)$ within $\Gamma_1$,
to the large-$x$ behavior $a_{eff}^R(\infty,s)=c$, after entering $\Gamma_2$. This is visualized in  
the inset of Fig.~\ref{fig4a}, displaying the behavior of
$a_{eff}^R(x,s)$ obtained analytically from Eq.~(\ref{aeffanall}), with the forms of $f(x)$ and $h(x)$ which will be
introduced in Eqs.~(\ref{scalf1}) and~(\ref{813}).

The next step is to focus, separately, on the two regions $\Gamma_1$ and $\Gamma_2$.   

\subsubsection{$\Gamma_1$ region}

We make the assumption that, with the range of $s$ reached in the simulations and $x \simeq 1$, we are exploring
the lower boundary of $\Gamma_1$, just above the $s$ axis, where $R_2(x,s)$ is negligible with respect to $R_1(x,s)$ (see Fig.\ref{fig5}).
The consistency of this hypothesis will be checked {\it a posteriori}. Let us, then, concentrate on the structure of
$R_1(x,s)$.
On the basis of existing analytical and numerical results, as discussed in 
detail in Ref.~\cite{resteso}, the scaling function in Eq.~(\ref{scalr}) is expected to be of the general form
\be
f(x,t_0/s)= A\frac{x^{-\beta}}{(x-1+t_0/s)^\alpha}
\label{scalf1}
\ee
where there appears the dependence on a microscopic time $t_0$, regularizing $f(x,t_0/s)$ at $x=1$.
This extra dependence on $t_0/s$, for $s$ large enough is negligible if
$x > 1$, but becomes crucial at $x=1$, where it is at the root of the observed discontinuity in the effective exponent.
In order to see this, let us replace Eq.~(\ref{aeffanall}) with the more precise form
\be
1+a_{eff}^R(x,s)=(1+a) \left [ \frac{ 1 + \frac{1}{1+a}\tilde{f}(x,v)+ \frac {(1+c)}{(1+a)} s^{-(c-a)}\kappa(x,v)}
{1+ s^{-(c-a)}\kappa(x,v)} \right ]
\label{n0}
\ee
where $v=t_0/s$, $\kappa(x,v)=h(x)/f(x,v)$ and the new term
\be
\tilde{f}(x,v) = v \frac{\partial_v f(x,v)}{f(x,v)}
\label{n1}
\ee
comes from $t_0/s$ in Eq.~(\ref{scalf1}). A simple computation yields
\be
     \tilde{f}(x,v) = \left \{ \begin{array}{ll}
        -\alpha (1- \frac{x-1}{v})  \qquad $for$ \qquad x-1 \ll v  \\ 	
         -\alpha \frac{v}{x-1}   \qquad $for$ \qquad x-1 \gg v
        \end{array}
        \right .
        \label{n2}
\ee 
which shows that this term modifies the effective exponent only at $x \simeq 1$, while $\kappa(x,v)$ keeps the same properties
of $\kappa(x)$, being finite for any finite $x$, including $x=1$,  and diverging for $x \rightarrow \infty$. 
As a result, for $s$ large but finite, we have
\be
    a_{eff}^R(x,s) = \left \{ \begin{array}{ll}
        a-\alpha  \qquad $for$ \qquad x=1   \\ 	
        a  \qquad $for$ \qquad 1 < x \ll \overline x (s) \\
        c  \qquad $for$ \qquad x \gg \overline x (s).
        \end{array}
        \right .
        \label{800}
\ee 
In the simulations, the behavior at $x=1$ can be studied with great precision. 
Considering the definition  $R_{ag}(1,s)=\lim _{\Delta s \to 0} R_{ag}(1+\Delta s/s,s)$,  
we have computed $R_{ag}(1,s)$ as $R_{ag}(1+\epsilon/s,s)$, 
using Eq.~(\ref{trebb}). This quantity is very easy to simulate, since good statistics can be obtained with
a modest numerical effort. The results are shown in
Fig.~\ref{fig6}, where, for completeness, we have included also the case $d=3$. 
The data show a neat algebraic decay, consistently with the assumption that $R_2(1,s)$
is negligible (if the correction was present, we  should have found  a crossover
between two different power laws). Relating the slope to the top line of Eq.~(\ref{800}), we find
\be
   1+a-\alpha  = \left \{ \begin{array}{ll}
        0.473\pm 0.001 \qquad $for$ \qquad d=2   \\ 	
        0.477\pm 0.001 \qquad $for$ \qquad d=3. 
        \end{array}
        \right .
        \label{numbers}
\ee 
These numbers need a comment. For such a short time difference 
$\chi \sim R$ and the Barrat conjecture leading to Eq.~(\ref{2.4}) is
correct, i.e. on the very short time regime the response function just mirrors the density of defects.
Indeed, we have measured independently the density of defects $\rho(s) \sim L^{-1}(s) \sim s^{-1/z}$, finding 
\be
   1/z  = \left \{ \begin{array}{ll}
        0.474\pm 0.001 \qquad $for$ \qquad d=2   \\ 	
        0.476\pm 0.001 \qquad $for$ \qquad d=3. 
        \end{array}
        \right .
        \label{numbers*}
\ee 
Therefore, the comparison of Eqs.~(\ref{numbers}) and~(\ref{numbers*}) leads to the identification 
\be
1/z = 1+a-\alpha
\label{alp}
\ee
which is, in fact, what one finds in the exact solution of  
the $d=1$ Ising model~\cite{Lippiello2000,Godreche2000,resteso} and in the approximate analytical
results at higher dimensionality~\cite{epj,Berthier,Mazenko04}. Analytical theories 
based on local scale invariance~\cite{Henkel2001}, instead, 
yield $\alpha = a +1$, in disagreement with
our data and with the aforementioned analytical results~\cite{nota+}.  

Once the negligibility of $R_2(x,s)$ in the $x\simeq 1$ region is established, the road to the direct measurement of $a$
is open. This is done rewriting Eqs.~(\ref{scalr}) and~(\ref {scalf1}), for $(t-s)/s \ll 1$, in the form
\be
s^{1+a-\alpha}R_1(x,s) = A(t-s+t_0)^{-\alpha}
\label{802}
\ee
which predicts, using Eq.~(\ref {alp}), that in the plot of $s^{1/z}R(x,s)$ versus $t-s$, the curves generated
for different $s$ ought to collapse, as long as $(t-s)/s \ll 1$ is satisfied. We have made such a plot (Fig.\ref{fig7}) of the data from the simulations
and using the value of $z$ in Eq.~(\ref{numbers*}) we have found  very good collapse for $(t-s) \leq 70$. 
From the slope of the curve in the region where there is collapse, we obtain $\alpha = 0.800 \pm 0.005$, from which, 
via Eqs.~(\ref {alp}) and~(\ref{numbers*}),
we get 
\be
a= 0.273 \pm  0.006
\label{803}
\ee 
which is close enough to $1/4$ to lend strong support to Eq.~(\ref{aphen}). 
We emphasise that this result, which has been obtained from the direct measurement of $R_{ag}(t,s)$,
reproduces the previous numerical results obtained for $a_{\chi}$ from ZFC~\cite{epj,resteso,Castellano2004,first}.
This is a nice check on the validity of Eq.~(\ref{2.5}), which disproves the recent claim~\cite{HPPA} 
that the exponent $a_{\chi}$ is unrelated to $a$. From the same set of data we have obtained also $A=0.153 \pm 0.002 J^{-1}$.

\subsubsection{$\Gamma_2$ region}

Clearly, to enter with the simulations so deep into $\Gamma_2$ 
that $R_1(x,s)$ can be neglected, is much more difficult. Nonetheless, the following
considerations are in order. 
Eq.~(\ref{vagiu}) requires that, for a given $s$, $R_{ag}(x,s)$ decays as
$x^{-\lambda _R/z}$ in this sector. Therefore,
the negligibility of $R_1(x,s)$ leads to the
remarkable consequence that this power law behavior
is a property of the correction $R_2(x,s)$ and not
of the leading  term $R_1(x,s)$. 
The implication is that the large $x$ behaviors of $h(x)$ and $f(x)$ are given by
\be
h(x) \sim x^{-\lambda _R/z}
\label{810}
\ee
\be
f(x) \sim x^{-(\alpha +\beta)}
\label{811}
\ee
with
\be
\alpha +\beta > \lambda _R/z.
\label{812}
\ee
This cannot be accounted for by local scale invariance~\cite{Henkel2001},
which assumes that the scaling function $f(x)$ of the dominant 
term decays like $x^{-\lambda _R/z}$ for $x \gg 1$.

\subsection{Fitting $R_{ag}(x,s)$} \label{fit}

\noindent Let us recapitulate what we have done so far, specifying the assumptions that we have introduced:

\begin{itemize}

\item the failure of simple scaling, displayed in Fig.~\ref{fig3}, requires to introduce a correction term.
We have {\it assumed} this to have the form of the correction to scaling appearing in Eq.~(\ref{2scalr}), with $c>a$.
This is enough to derive the inequality of Eq.~(\ref{exponent}), which solves the problem of the exponent $a$
formulated in Sec.~\ref{sectexponent}.

\item From the two requirements, $a_{eff}^R(1^+,s) < a_{eff}^R(\infty,s) < \infty$ and $\lim_{x \rightarrow \infty}R_{ag}(x,s) = 0$, 
follows $a_{eff}^R(\infty,s) = c$ and  $\lim_{x \rightarrow \infty}\kappa(x) = \infty$. This, together with
the {\it assumption} that $f(x)$ is of the form~(\ref{scalf1}), yields the value~(\ref{803}) of $a$ and 
the set of Eqs.~(\ref{810},\ref{811},\ref{812}) regulating the large $x$ behavior of the scaling functions.

\end{itemize}

\noindent In order to complete the functional form of $R_{ag}(x,s)$, we need an additional assumption
on $R_2(x,s)$, since the power law behavior~(\ref{810}) covers only the large $x$ behavior.
However, for $R_2(x,s)$ there are no analytical results to rely on. Short of any other hint, we make the simplest possible ansatz
\be
h(x)=B x^{-\lambda _R/z}
\label{813}
\ee
which continues the power law behavior~(\ref{810}) also into the short time region $x \simeq 1$.
Then, we arrive to the explicit analytical expression
\be
R_{ag}(x,s)= As^{-(1+a)}\frac{x^{-\beta}}{(x-1+v)^\alpha}+
Bs^{-(1+c)}x^{-\lambda _R/z}
\label{fitting}
\ee
for $R_{ag}(x,s)$ in the $d=2$ Ising model.
We do not expect this to be quantitatively exact, rather with this formula we aim to capture the gross
features of the qualitative behavior. In any case, the evaluation of the performance of this formula has to
be made {\it a posteriori}.
The program is i) to fit the remaining free parameters in Eq.~(\ref{fitting}) from the data, ii) to check how 
the predicted values for $R_{ag}(x,s)$ compare with the numerical ones and iii) to extrapolate to the region of $x$ and $s$ that
have not been reached with the simulations (see Fig.\ref{fig5}).

\begin{figure}
    \centering
   \rotatebox{0}{\resizebox{1\textwidth}{!}{\includegraphics{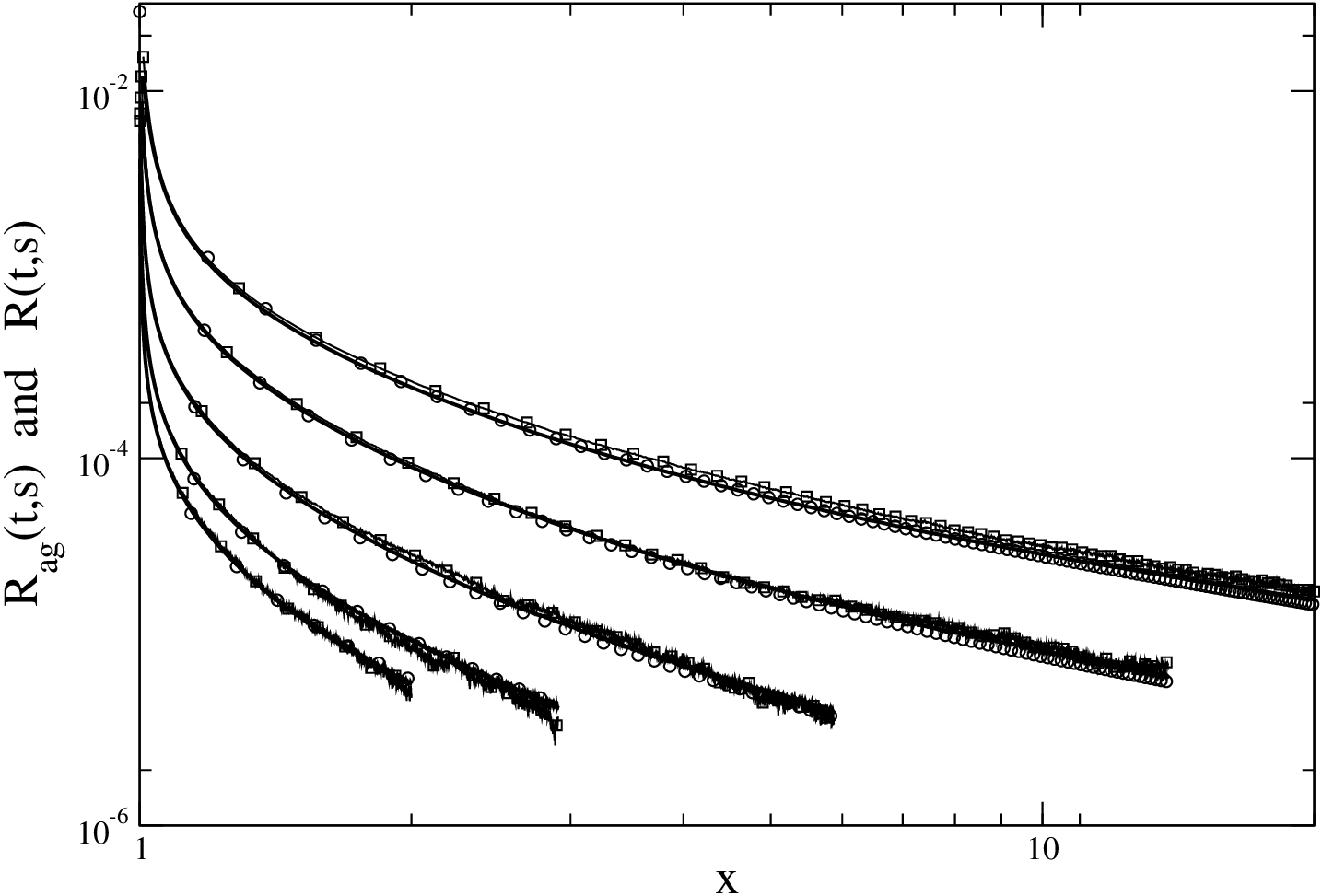}}}
    \caption{$R_{ag}(x,s)$ (continuous lines) obtained with the NBF algorithm and 
      $R(x,s)$ (squares)  obtained with the full quench dynamics  are  plotted 
      for $s=111,229,537,1085,1577$ from top to bottom. 
      Circles represent the fit according to Eq.~(\ref{fitting}).} 
\label{fig2}
\end{figure}

Our data do not allow a direct precise determination of $\lambda _R$ because,
as Fig.~\ref{fig2} shows, the slope of $R_{ag}(x,s)$ bears a weak dependence on $x$ 
even for the largest values of $x$ reached in the simulations.
Therefore, we take $\lambda _R=\lambda$ and $\lambda=1.25$~\cite{Mazenko91}
which is consistent, within errors, with the data. 
For the other parameters $t_0,\beta,B,c$ 
we have not found a reliable direct method to measure them. We have used a four-parameter fitting routine
obtaining $B= 0.47 J^{-1}$, $c=0.81$, $t_0= 0.01$, 
$\beta = 1.1$.
With these parameters, and ($a,\alpha,A$) determined in the previous section~\ref{erre1},
Eq.~(\ref{fitting}) can be plotted. The comparison with the numerical data (Fig.~\ref{fig2})
is quite good in the whole range of 
$s$ and $x$. This provides the {\it a posteriori} support for the validity of the procedure.

Once $f(x,v)$ and $h(x)$ are given, from  
Eq.~(\ref{n0}) we get the analytical form of $a_{eff}^R(x,s)$, which can be compared with the data of
Figs.~\ref{fig4a} and~\ref{fig4b}. However, for a
meaningful comparison, rather than making a straightforward plot of
Eq.~(\ref{n0}), we have extracted the effective exponent from data generated from Eq.~(\ref{fitting}), 
following the same procedure described in Sec.~\ref{effect}.
Namely, we have computed $R_{ag}(x,s)$, from 
Eq.~(\ref{fitting}), for the same values
of $s$ and $t$ considered in the simulations and we have
computed $a^R_{eff}(x,s)$ as the local slope over the
same intervals $I_s$ of four values of $s$ appearing in Fig.~\ref{fig4a}. 
The result is displayed in the same Figs.~\ref{fig4a} and~\ref{fig4b}, where, although
a discrete set, these values of $a_{eff}^R(x,s)$ have been connected by continous lines,
in order to ease the reading of the Figures. The comparison shows
that the behavior of $a_{eff}^R(x,s)$ from Eq.~(\ref{fitting}) reproduces the numerical data
within errors. However, the size of the error bars prevent from making claims other than qualitative.
Namely, the agreement displayed in Figs.~\ref{fig4a}  and~\ref{fig4b}, between the different ways of computing
$a_{eff}^R(x,s)$, makes us confident that with Eq.~(\ref{fitting}) we have captured the basic
mechanism underlying the behavior of $R_{ag}(x,s)$ in the $d=2$ Ising model. For a detailed
quantitative comparison, data much more precise than those presented here are needed.

With the aim of completing the qualitative scenario,
we have used Eq.~(\ref{n0}) to plot $a_{eff}^R(x,s)$, in the inset of Fig.~\ref{fig4a}, over ranges of
$s$ and $x$ that cannot be accessed in the simulations.
The global behavior of $a_{eff}^R(x,s)$, obtained in this way, displays the typical shape of an
effective exponent with the crossover taking place about 
$\overline x(s)$. For $s$ sufficiently large, $a_{eff}^R(x,s)$ remains equal to
$a$ in the range $1<x\ll \overline x(s)$, which can be enlarged at will by pushing
$s$ to larger values. The crossover around $\overline x(s)$ shows 
the rise of $a_{eff}^R(x,s)$ toward $c$.
It is a very smooth crossover, since it takes roughly three decades
for $a_{eff}^R(x,s)$ to switch from $a$ to $c$. The growth of $\overline x(s)$ with $s$
is rather slow and a value as large as $s\simeq 10^4$
is needed in order to have a reasonably flat behavior $a_{eff}^R(x,s)\simeq a$ for $x < \overline x(s)$.
This explains why the observation of $a_{eff}^R(x,s)\simeq a$ over a sizable $x$ interval is out of reach
in the simulations.

\section{Integrated response functions} \label{irf}

In Eq.~(\ref{irff}) we have introduced the general form of the IRF, which contains ZFC~(\ref{0.1})
as a particular case. The aim of this section is to investigate the impact
on the IRF of the structure~(\ref{2scalr}) of $R_{ag}(x,s)$. In particular, we shall have to understand
why the existence of such a strong correction to scaling in $R_{ag}(x,s)$ does not surface
at the level of ZFC, where the simple scaling form~(\ref{2.1}) accounts very well for the data~\cite{resteso}.
We shall not discuss the behavior of TRM 
because, as mentioned above and explained in detail in~\cite{resteso,Comment},
this quantity is the most unfavorable choice for the analysis of the scaling properties.

Inserting Eq.~(\ref{2scalr}) into the definition~(\ref{irff}) and keeping track of
the dependence on $t_0$, we get
\be
\mu _{ag}(y,t_w)=\mu _1(y,t_w)+\mu _2(y,t_w)=
t_w^{-a}F(y,v)+t_w^{-c}H(y)
\label{omega2}
\ee
where 
\be
F(y,v) = y^{-a} \int _{1/y}^u z^{-(1+a)} \phi (z,v/y) dz
\label{omega2.1}
\ee
\be
H(y) = y^{-c} \int _{1/y}^u z^{-(1+c)} h(1/z) dz
\label{omega2.2}
\ee
with $y=t/t_w,u=\tilde{t}/t,v=t_0/t_w$ and 
\be
\phi (z,v/y) = A{z^{\alpha + \beta} \over (1-z +v/y)^\alpha}
\label{2scalmu.3}
\ee
comes from Eq.~(\ref{scalf1}) substituting $x$ with $1/z$.
For simplicity, we have omitted to indicate explicitely the dependence on the upper limit of integration $u$.
The effective exponent 
\be
a_{eff}^\mu(y,t_w) =  - \left. {\partial \ln \mu _{ag} (y,t_w)} 
\over {\partial \ln t_w} \right |_{y} 
=  a \left [ \frac{ 1 + \frac{1}{a}\widetilde{F}(y,v)+ \frac {c}{a} t_w^{-(c-a)} K(y,v)}
{1+ t_w^{-(c-a)} K(y,v)} \right ]
\label{300}
\ee
has the same structure of Eq.~(\ref{n0}), with
\be
\widetilde{F}(y,v) = v{\partial_v F(y,v) \over F(y,v)}
\label{711}
\ee
and
\be
K(y,v) = {H(y) \over F(y,v)}
\label{71}
\ee
being the analogues of $\tilde{f}(x,v)$ and $\kappa(x,v)$.

There are as many IRF as there are ways of choosing $\tilde{t}$.
Here, we restrict the attention to the two cases with $\tilde{t}/t_w=q$
and  $\tilde{t}/t=p$, where $q$ and $p$ are fixed numbers. It will be convenient to introduce the notation
$\omega_{ag}$ and $\pi_{ag}$ for these two particular IRF. The distinction between the two cases enters the above formulas
in the upper limit of integration in Eqs.~(\ref{omega2.1}) and~(\ref{omega2.2}), where 
\be
   u  = \left \{ \begin{array}{ll}
        q/y \qquad $for$ \qquad \omega_{ag}  \\ 	
        p \qquad $for$ \qquad \pi_{ag}. 
        \end{array}
        \right .
        \label{730}
\ee 
We shall now establish the gross features of $a_{eff}^\mu(y,t_w)$, as $y$ is varied with fixed $t_w$, using the properties of
$\widetilde{F}(y,v)$ and $K(y,v)$ derived in the Appendix. The behavior of these functions depends on the sign of $\alpha-1$.
With the value of $\alpha=0.80$ obtained from the fit of Eq.~(\ref{802}), we shall make use only of the results 
for $\alpha < 1$, leaving the general discussion to the Appendix.

\subsection{ $\mathbf \omega_{ag}$  } \label{V-A}

With $y \geq \tilde{t}/t_w=q$, where $q >1$ is some fixed number, $y$ takes values in $[q,\infty)$. 
From Eqs.~(\ref{a133}) and~(\ref{a135}) in the Appendix follows
\be
    \widetilde {F}_{\omega}(y,v)  \sim \left \{ \begin{array}{ll}
         (v/q)^{1-\alpha} \qquad $for$ \qquad y=q  \\ 	
        (v/y) \qquad $for$ \qquad y > q 
        \end{array}
        \right .
        \label{m0}
\ee 
while, from Eqs.~(\ref{a136}) and~(\ref{a138tris}) follows that $K_{\omega}(y,v)$
is finite for finite $y$ and diverges for $y \rightarrow \infty$. Hence,  the affective exponent
\be
   a_{eff}^\omega(y,t_w) = \left \{ \begin{array}{ll}
        a \qquad $for$ \qquad y \simeq q  \\ 	
        c \qquad $for$ \qquad y \rightarrow \infty
        \end{array}
        \right .
        \label{m1}
\ee 
behaves like  $a_{eff}^R$ in Eq.~(\ref{800}), apart from the absence of any discontinuity at the minimum value of $y$, 
since Eq.~(\ref{m0}) shows that $\widetilde {F}_{\omega}(y,v)$
vanishes for all $y$ as $t_w \rightarrow \infty$. This is displayed in Fig.\ref{fig8} illustrating the behavior
of  $a_{eff}^\omega(y,t_w)$ obtained from the simulations with $t_w$ in the range $[200,400]$.

\subsection{ $\mathbf \pi_{ag}$} \label{V-B}

If we keep fixed the ratio $\tilde{t}/t = p$, with  $p \leq 1$, $y$ takes values in $[1/p,\infty)$.
In this case we obtain an IRF 
in which the correction to scaling does not produce the crossover in the effective exponent from
$a$ to $c$, as observed in the previous case and in $R_{ag}$. Rather, $a_{eff}^\pi(y,t_w)$ starts from $a$
and ends up again to $a$, as $y$ grows, with a possible discontinuity at $y=1/p$, as
will be explained below.
The difference comes from Eq.~(\ref{a43}) in the Appendix, showing that now $K_\pi(y,v)$ vanishes
as $y \rightarrow \infty$. After taking into account that also $\widetilde {F}_\pi(y,v)$ vanishes as  $y \rightarrow \infty$     
(see  Eq.~(\ref{a4}) in the Appendix), this implies
\be
\lim_{y \to \infty}  a_{eff}^\pi(y,t_w) = a.
\label{m2}
\ee
Around the minimum value $y=1/p$, the behavior is different for $p<1$ and $p=1$.
In the former case no singularity develops, since from Eq.~(\ref{a41}) follows that $\widetilde {F}_\pi(y=1/p,v) \simeq v$. 
Instead in the latter case,  which corresponds to ZFC (i.e. $\pi_{ag}=\chi_{ag}$ for $p=1$), from Eq.~(\ref{a45}) we have 
that at $y=1$ there is a discontinuity in the effective exponent with
\be
a_{eff}^\chi(1,v,t_w) =  a-\alpha
\label{m3}
\ee 
exactly as in Eq.~(\ref{800}) at $x=1$.

The overall behavior of $a_{eff}^\chi(y,t_w)$ is displayed in Fig.\ref{fig9} obtained by plotting Eq.~(\ref{300}) for fixed $t_w$.
The two curves, corresponding to $t_w=200$ and $t_w=2000$, display a very fast rise from $a-\alpha=-0.53$ at $y=1$ followed
by a very slow approach to the asymptotic value~(\ref{m2}). The absence of the crossover is the most prominent qualitative
difference with respect to the inset of Fig.\ref{fig4a}, which explains why the correction to scaling plays a minor role in
the analysis of ZFC data. This very mild increase of $a_{eff}^\chi(y,t_w)$ in the range $1 < y \leq 10$ is observed
also in the simulations (notice the vertical scale in Fig.\ref{fig10}).

Furthermore, notice that in Fig.\ref{fig9} the  asymptotic value is reached
from above for $t_w=200$ and from below for $t_w=2000$. The two curves cross each other,
due to the interplay of the relative weights of $\widetilde{F}_{\chi}(y,v)$ and $K_{\chi}(y,v)$ as $t_w$ is varied,
the former being a negative quantity and the latter a positive one.
Now, the important point is that this particular feature can be resolved also in the data
from the simulations (Fig.~\ref{fig10}), showing that Eq.~(\ref{fitting}) does, indeed, account for the observed 
behavior of ZFC.

\section{Conclusions} \label{concl}

In this paper we have studied numerically the linear response function of the
$d=2$ Ising model quenched below the critical temperature. The data for
$R_{ag}$ show that the simple scaling form~(\ref{scalr}), usually believed to hold in
the aging regime, is not obeyed. We have attributed the deviation from
simple scaling to the existence of a correction to scaling $R_2(x,s)$,
which, in order to explain the observed behavior of the effective exponent, although
subdominant for fixed $x$ and growing $s$, must become dominant for fixed $s$ and
large $x$. Then, focusing on the time regime where  $R_2(x,s)$ is negligible, we have
been able to measure $a$ with good precision obtaining $a=0.273 \pm 0.006$.
This solves the problem of the exponent $a$ formulated in Sec.~\ref{sectexponent} and 
is in agreement with previous numerical results from the
ZFC. Furthermore, this value is well consistent with Eq.~(\ref{aphen}) and confirms
the dependence of $a$ on dimensionality, as found in all cases
where analytical results are available. 

\begin{figure}
    \centering
   \rotatebox{0}{\resizebox{1\textwidth}{!}{\includegraphics{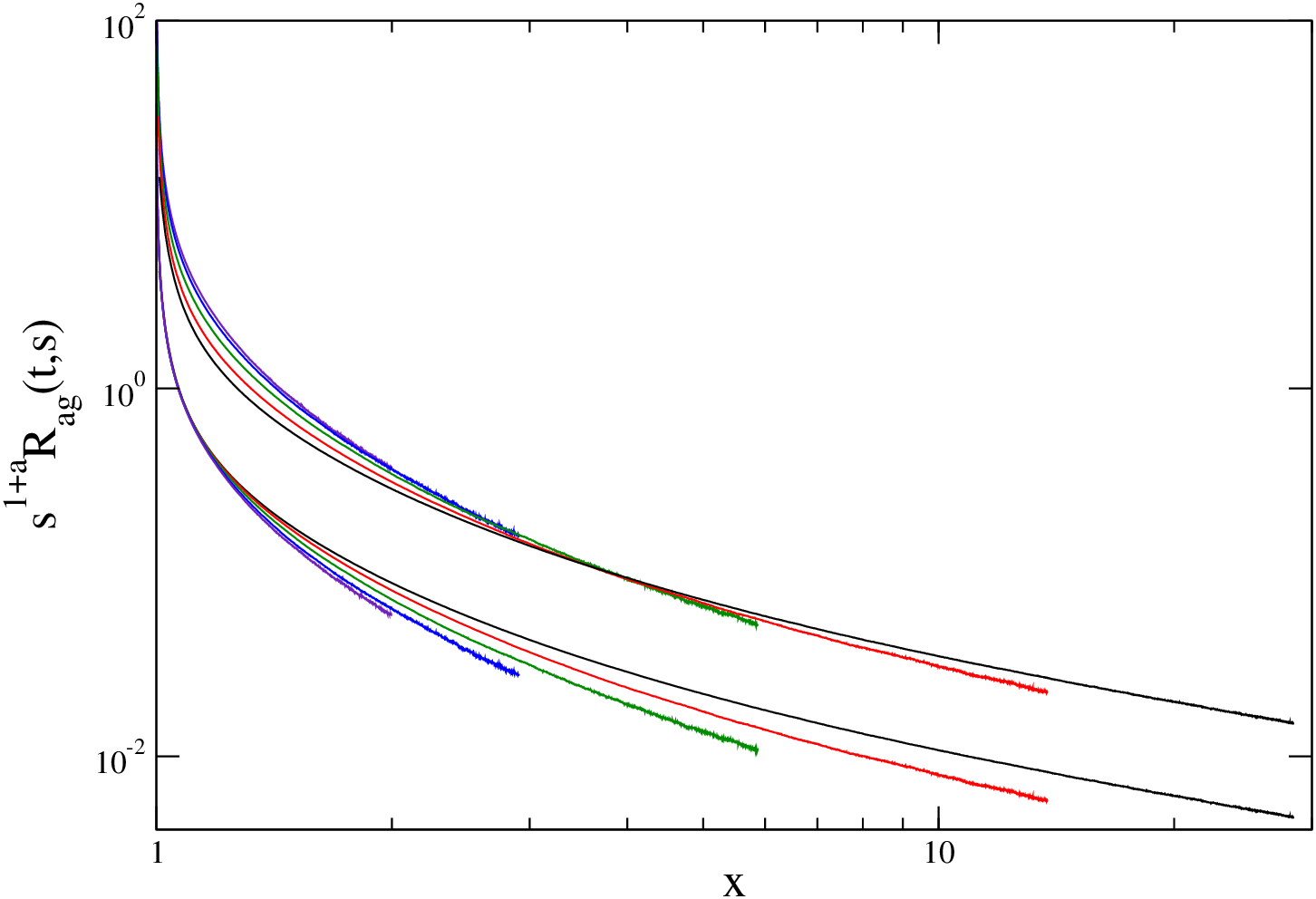}}}
    \caption{Failure of collapse of the curves of Fig.~\ref{fig2} with
     $a=1/4$ (lower set) and $a=1/2$
    (upper set).}
\label{fig3}
\end{figure}

\begin{figure}
    \centering
   \rotatebox{0}{\resizebox{1\textwidth}{!}{\includegraphics{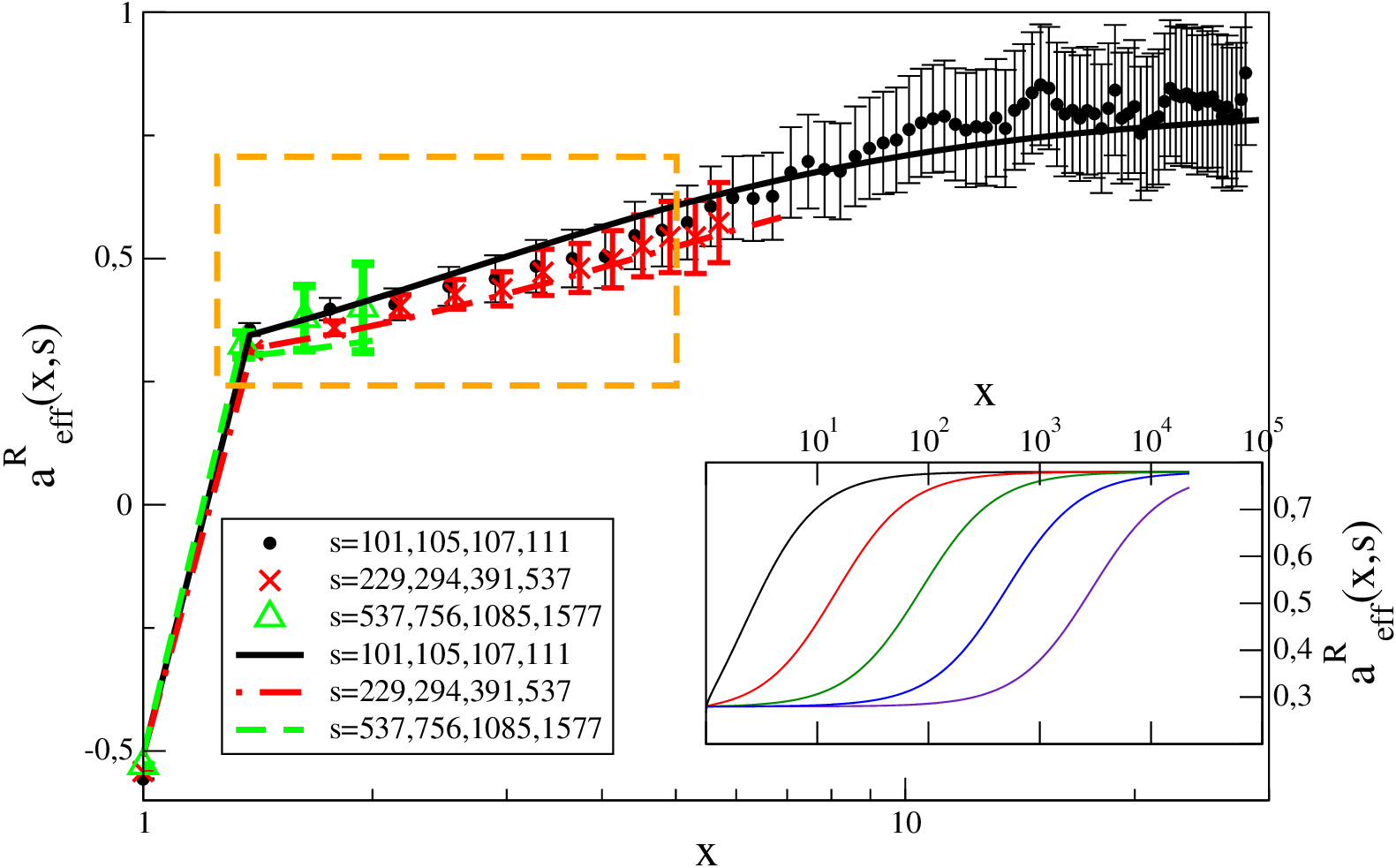}}}
    \caption{ 
Symbols with error bars represent $a^R_{eff}(x,s)$ obtained from the simulation data,  as  local slopes 
of $\ln R (x,s)$ against
$\ln s$ over the four values of $s$ specified in the legend. Continuous curves are
the plot of $a^R_{eff}(x,s)$ obtained from the fitting 
formula~(\ref{fitting})
for the same $s$ used in the simulations. The frame is magnified in Fig.~\ref{fig4b}.
The inset shows $a^R_{eff}(x,s)$ from 
formula~(\ref{fitting}) over a range $s=3\cdot 10^2, 10^3,10^4,10^5,10^6$ (from top to bottom) 
that cannot be reached in the simulation.}
\label{fig4a}
\end{figure}

\begin{figure}
    \centering
   \rotatebox{0}{\resizebox{1\textwidth}{!}{\includegraphics{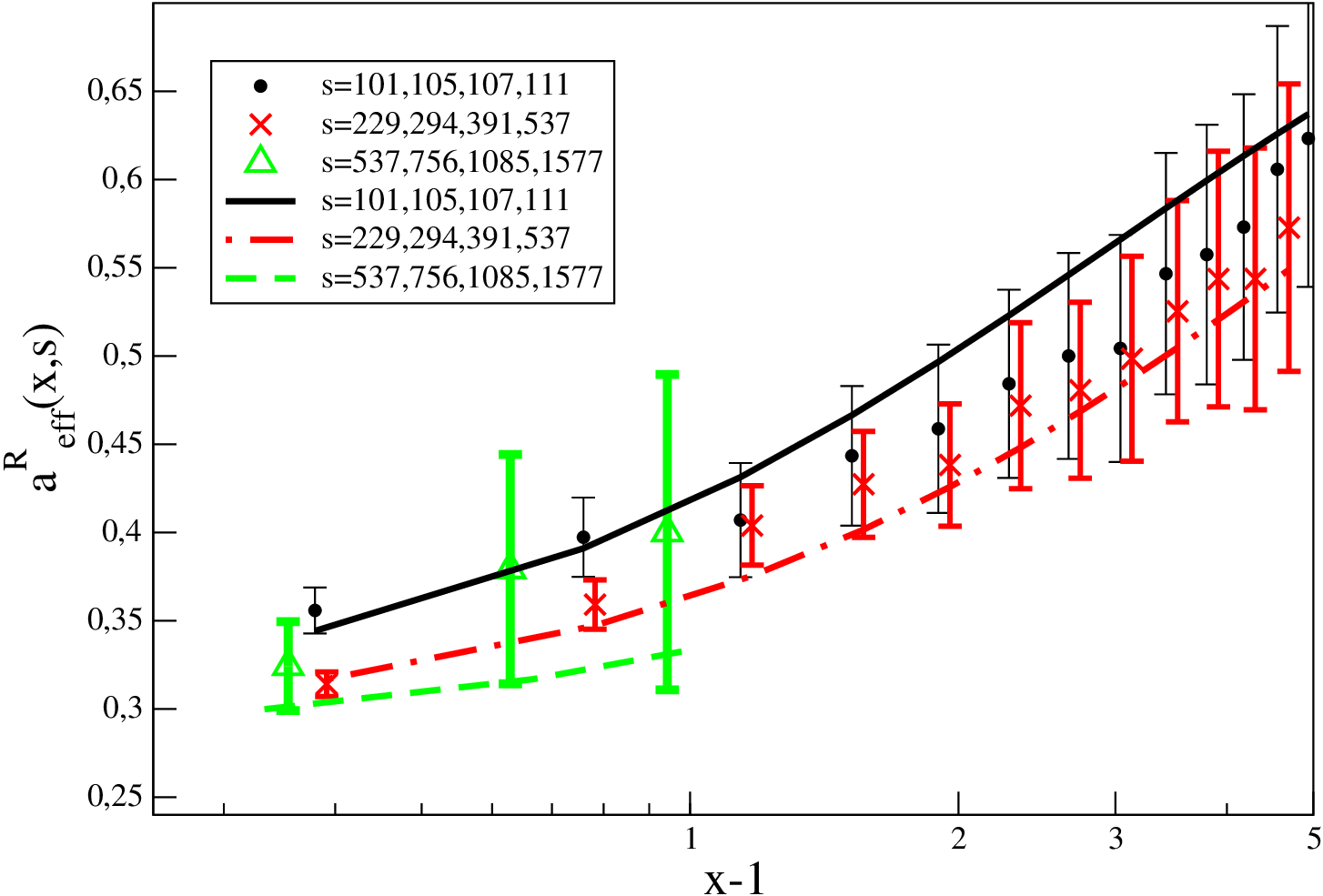}}}
    \caption{Magnification of the framed portion of Fig.~\ref{fig4a}, showing that
    the decrease of $a^R_{eff}(x,s)$ with increasing $s$ (limited to the first two sets $I_s$) 
    exceeds the error bars for  $x \leq 2$. Symbols are the same as in Fig.~\ref{fig4a}.}
\label{fig4b}
\end{figure}

\begin{figure}
    \centering
   \rotatebox{0}{\resizebox{1\textwidth}{!}{\includegraphics{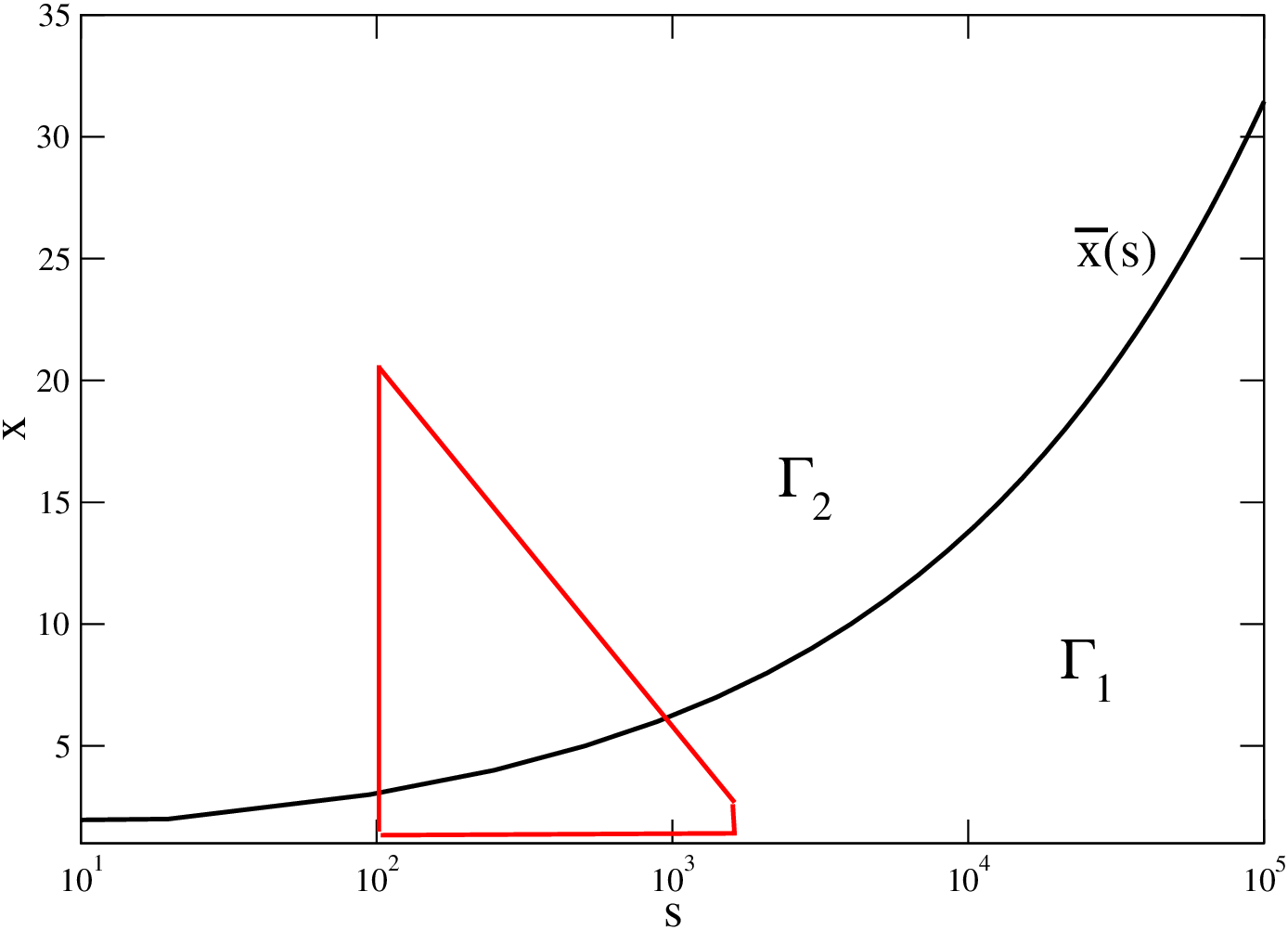}}}
    \caption{
    Shape of the crossover curve computed from Eq.~(\ref{ratio}). The frame delimits the $(x,s)$ 
    region explored in the simulations.}
 
\label{fig5}
\end{figure}

\begin{figure}
    \centering
   \rotatebox{0}{\resizebox{1\textwidth}{!}{\includegraphics{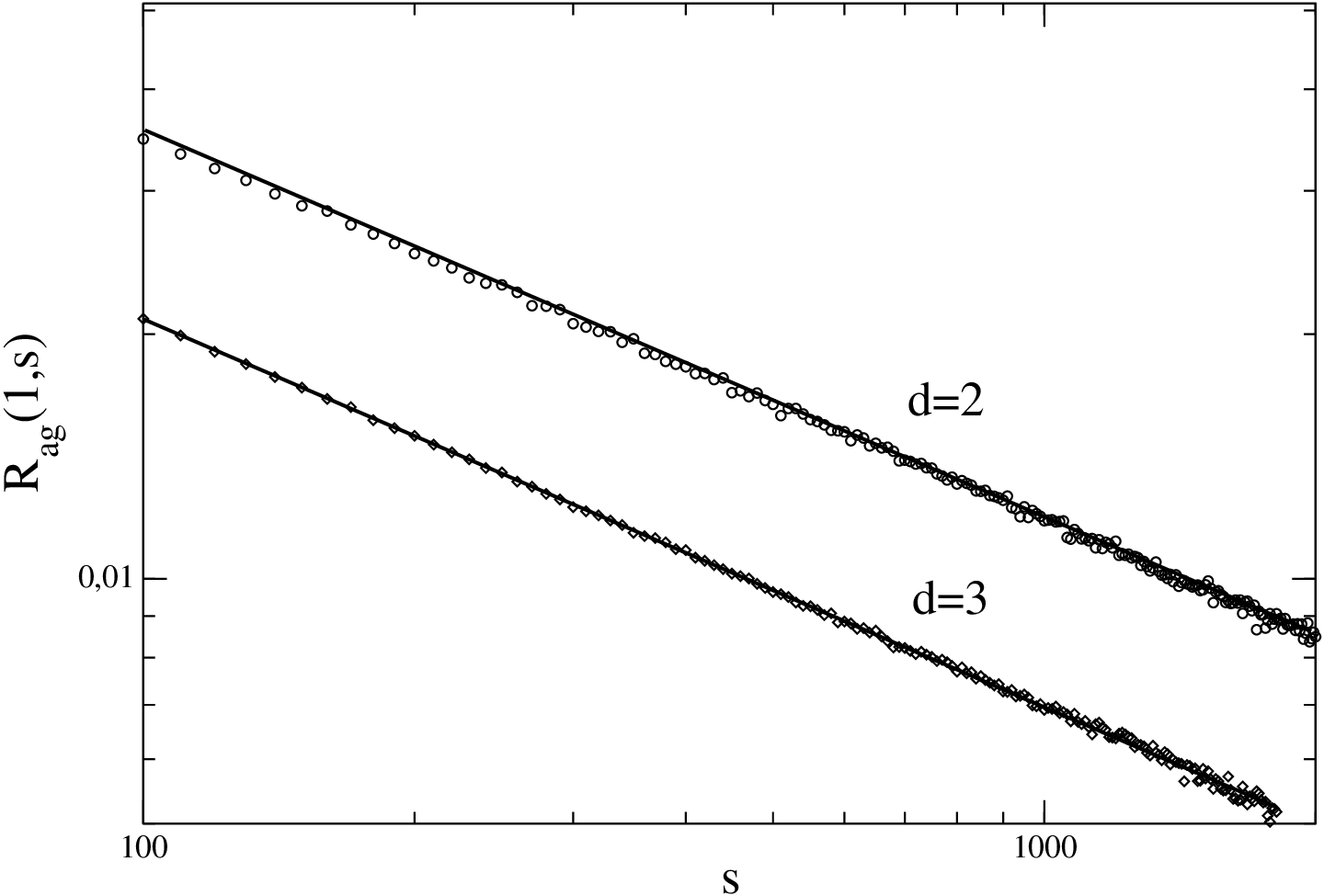}}}
    \caption{Plot of $R_{ag}(1,s)$ in $d=2$ and $d=3$. Straight lines
    are power law best fits $t^{-0.473}$ in $d=2$ and $t^{-0.477}$ in $d=3$.}
 \label{fig6}
\end{figure}

\begin{figure}
    \centering
   \rotatebox{0}{\resizebox{1\textwidth}{!}{\includegraphics{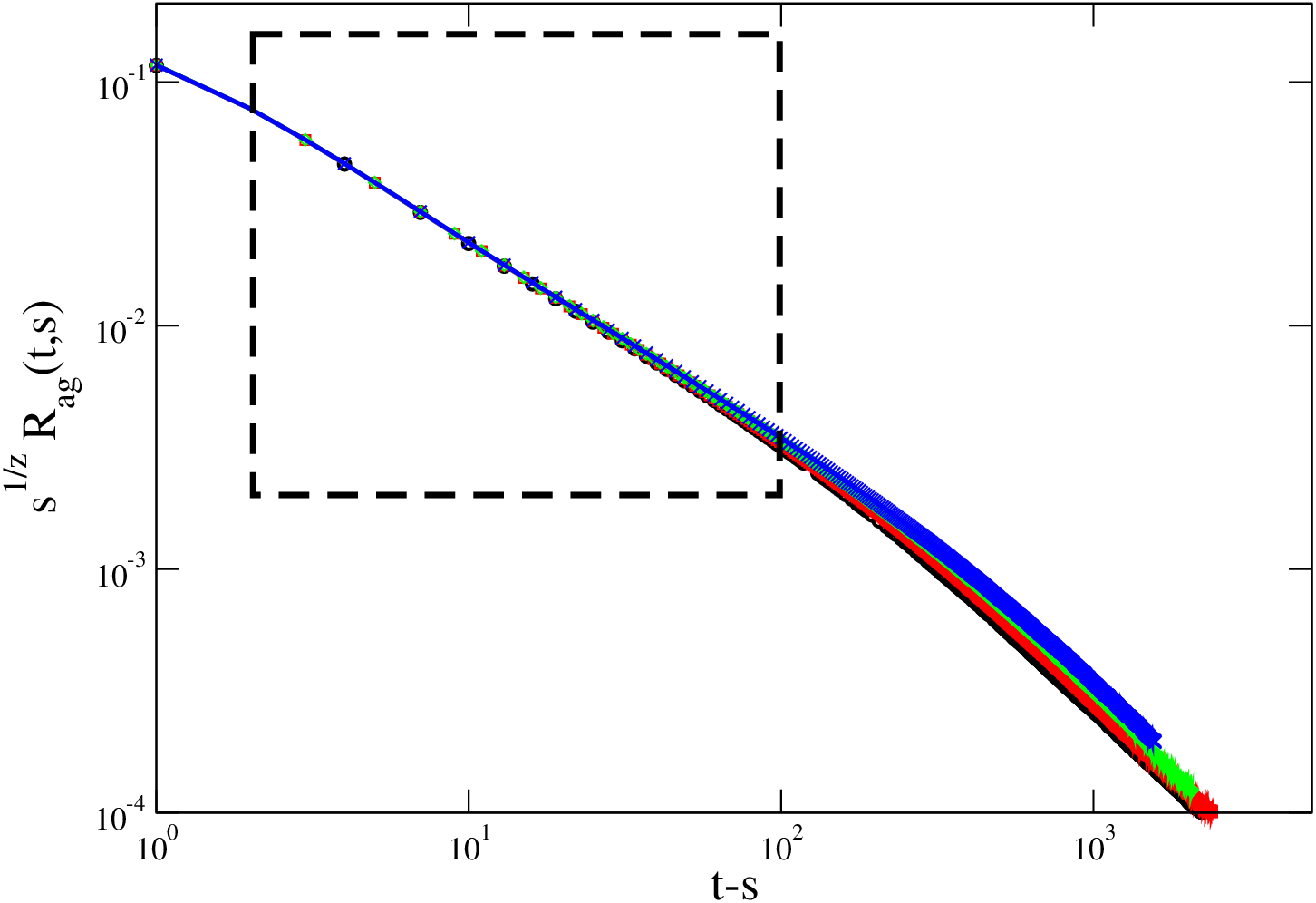}}}
    \caption{
            Rescaled plot of $R_{ag}$ versus $t-s$ for $s=537,756,1085,1577$. Power law best fit in
            the framed area $(t-s)^{-0.80}$.}
 \label{fig7}
\end{figure}

\begin{figure}
    \centering
   \rotatebox{0}{\resizebox{1\textwidth}{!}{\includegraphics{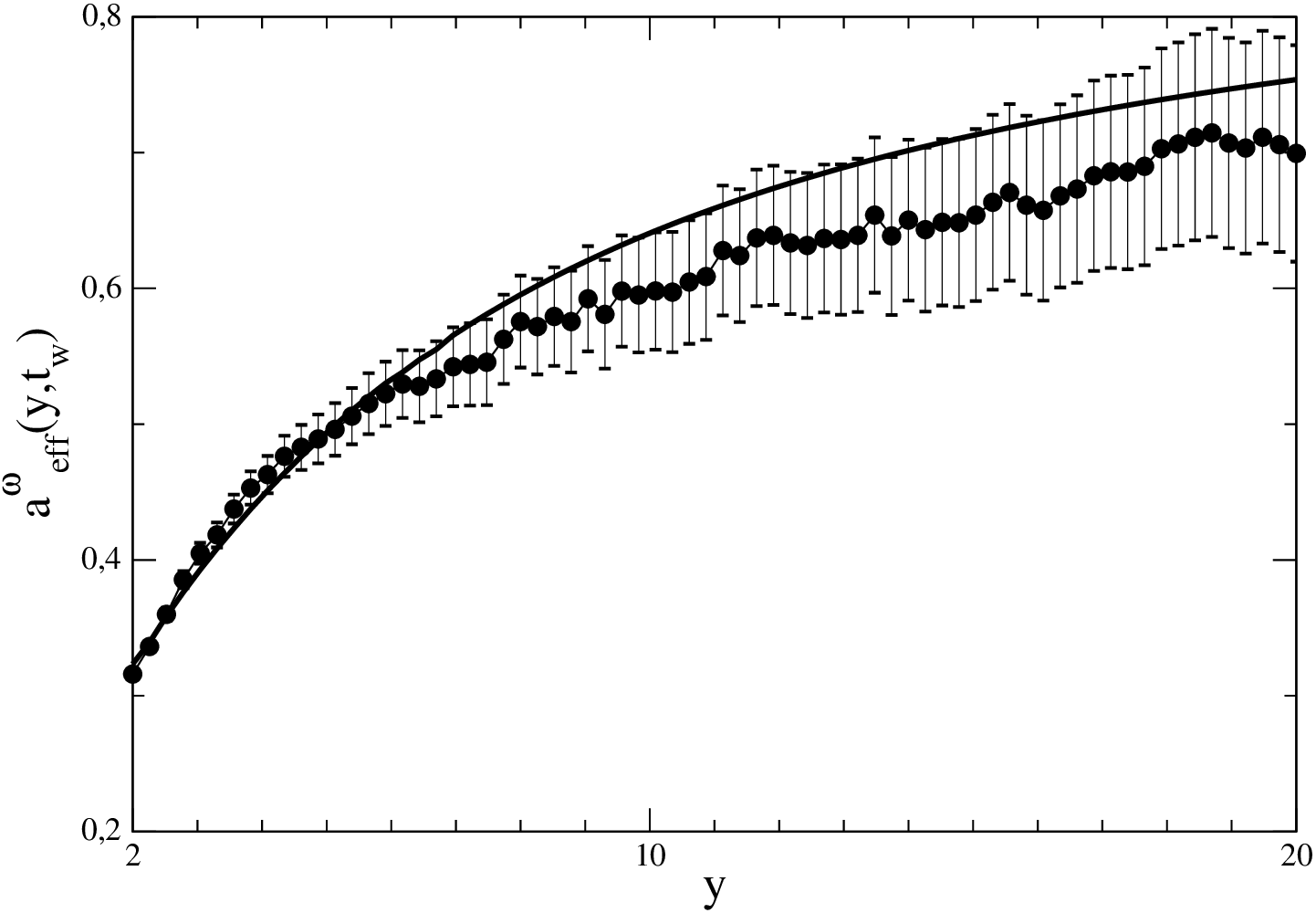}}}
    \caption{
            $\alpha^{\omega}_{eff}(y,t_w)$ for $t_w$ in the range from $200$ to  $400$. The continous line is the
            plot from Eq.~(\ref{300}).}
\label{fig8}
\end{figure}

\begin{figure}
    \centering
   \rotatebox{0}{\resizebox{1\textwidth}{!}{\includegraphics{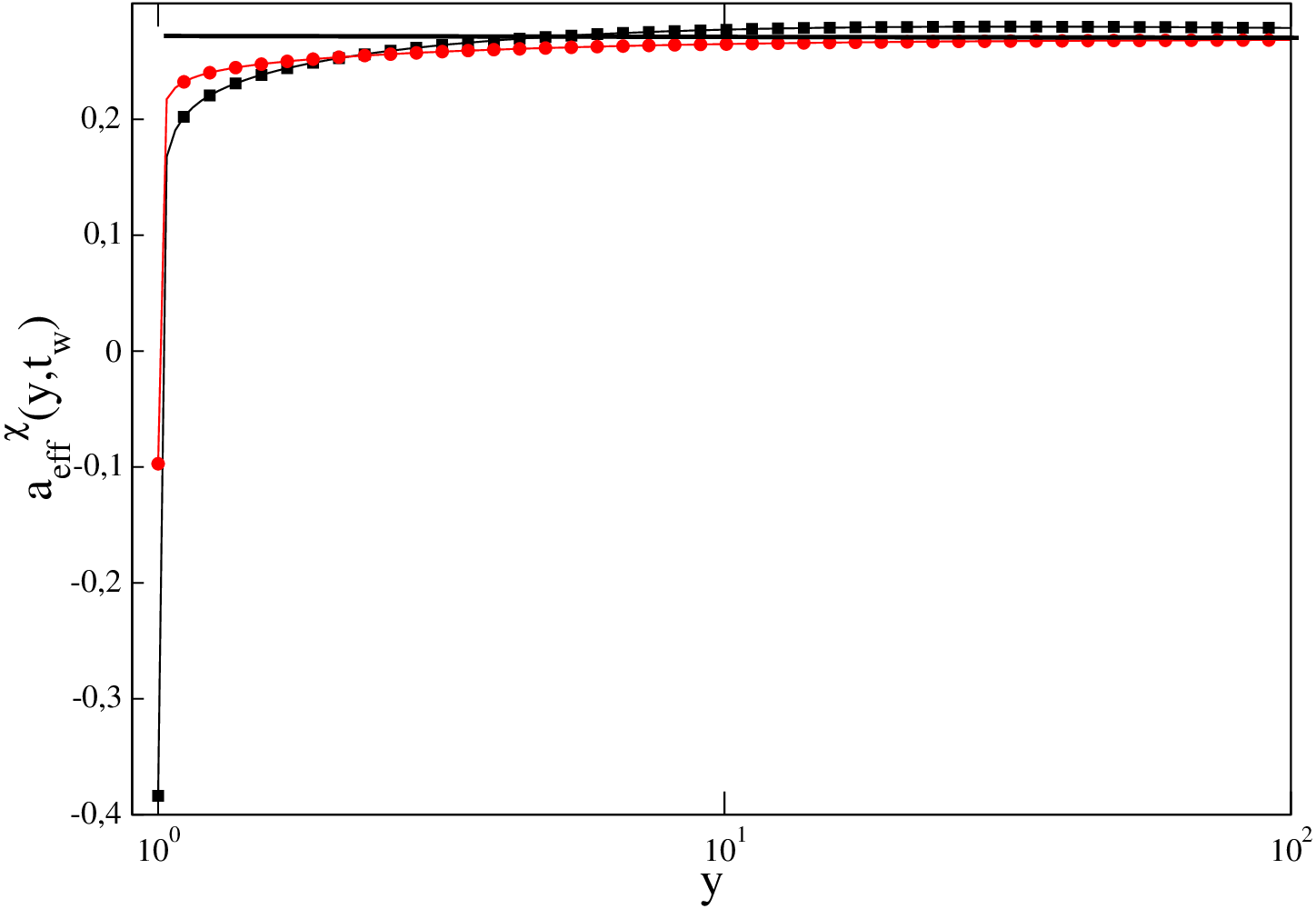}}}
    \caption{$\alpha^{\chi}_{eff}(y,t_w)$ plotted from Eq.~(\ref{300})with $t_w=200$ (squares) and $t_w=2000$ (circles).
    The horizontal line corresponds to $a=0.273$.}
 
\label{fig9}
\end{figure}

\begin{figure}
    \centering
   \rotatebox{0}{\resizebox{1\textwidth}{!}{\includegraphics{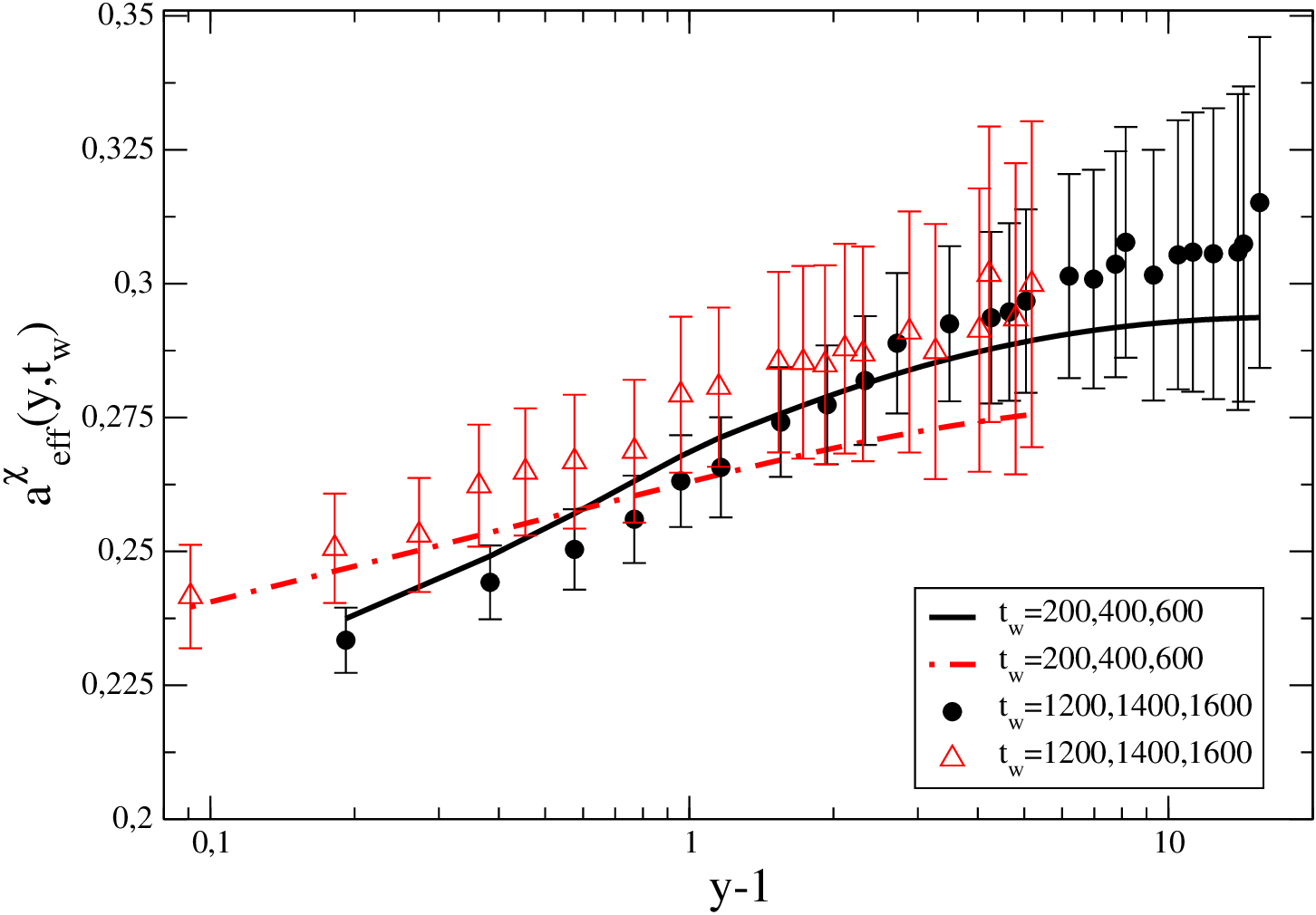}}}
    \caption{Plot of $a^{\chi}_{eff}(y,t_w)$ from the simulations for the two sets of $t_w$
             in the legend. The lines are obtained from the analytical formula.}
 
\label{fig10}
\end{figure}

Awareness of the existence of this correction to scaling is of fundamental
importance when analyzing numerical data, since an interpretation which 
does not take into account the strong dependence of $a^R_{aff}(x,s)$ on $x$
may lead to wrong conclusions. In particular, if insisting in collapsing the
data according to Eq.~(\ref{scalr}), one would find that the 
best data collapse obtains with an exponent whose value is somewhere between 
$a$ and $c$, depending on the range of $x$ considered in the simulations.
However, the collapse cannot be satisfactory in the whole range of $x$, as Fig.~\ref{fig3} shows.
The findings of Chatelain~\cite{Chatelain03}, perhaps,  can be interpreted along
these lines. Indeed, in Ref.~\cite{Chatelain03} the impulsive response function was
computed and the best data collapse was obtained with $a=1/2$ using data for rather small values
of $s$. The author did explicitly notice that the collapse was not perfect
and hypothesized himself the existence of strong corrections to scaling.

We have also investigated the functional form of the scaling function entering $R_1(x,s)$, 
and we have found that Eq.~(\ref{scalf1}) 
with $\alpha=a+1/z$, which was proposed in~\cite{resteso} as a generalization of the  known 
analytical results, compares quite well with the numerical data.
Next, by making the simplest possible ansatz for $R_2(x,s)$, we have obtained
an analytical expression for the full $R_{ag}$. This is quite useful for
the investigation of the overall qualitative behavior and compares satisfactorily
well with the data in the time regions reached by the simulations.

Finally, we have considered the retrival of
the properties of the impulsive response function from the IRF. 
As discussed in~\cite{resteso}, this may be an issue to treat with care when dealing
with ZFC, due to the presence of $t_0$, which acts as a dangerous irrelevant variable for $\alpha \geq 1$ or,
equivalently, for $ d \geq d_U$. However, in this paper we are not concerned with this aspect of the 
problem, since  $d=2$ is below $d_U=3$ and we have $\alpha=0.80 < 1$. Nonetheless,
recovering the scaling exponent may still be 
complicated, due to the presence of the correction $R_2(t,s)$.
In fact, we have found that the relative importance of this term 
strongly depends on the kind of response function considered. In particular, it is 
maximal in $R_{ag}(t,s)$, while it is almost negligible for
$\chi_{ag}(t,s)$. This implies that
i) by weighting differently $R_1(t,s)$ and $R_2(t,s)$, different
response functions may behave very differently and
ii) the best suited function
in order to weaken the correction term and to access the asymptotic
properties is $\chi_{ag}(t,s)$.  

The rich pattern of behaviors uncovered in
this paper shows that there remains much to be understood of the response function in slow relaxation
phenomena, even in the relatively simple 
case of coarsening systems. Presently, we do not know how general is the 
behavior that we have found in $d=2$. 
It would be interesting to perform simulations in
higher dimension to address this point, at least numerically.

\section{Appendix} \label{appendix}

In order to analyse the behavior of
\be
a_{eff}^\mu(y,t_w)
= a \left [ \frac{ 1 + \frac{1}{a}\widetilde{F}(y,v)+ \frac {c}{a} t_w^{-(c-a)} K(y,v)}
{1+ t_w^{-(c-a)} K(y,v)} \right ]
\label{a00}
\ee
it is first necessary to study the properties of $\widetilde {F}(y,v)$ and $K(y,v)$. Rewriting Eq.~(\ref{omega2.1}) as 
\be
F(y,v)= A y^{-a} I(\alpha) 
\label{a0}
\ee
with
\be
I(\alpha) = \int _{1/y}^u dz {z^{\beta + \alpha - (1+a)} \over (1-z +v/y)^\alpha} 
\label{a2}
\ee
we have
\be
\partial_v F(y,v) = -A \alpha  y^{-(1+a)} I(\alpha+1)
\label{a00b}
\ee
and
\be
\widetilde {F}(y,v) = -\alpha \left ( \frac{v}{y} \right ) { I(\alpha+1) \over I(\alpha)}.
\label{a4}
\ee
Next, carrying out the integration in~(\ref{omega2.2}), we can write
\be
K(y,v) =  {\cal B} {y^{a-c} \over  I(\alpha)} [u^{\lambda_R/z-c} - y^{c-\lambda_R/z}]
\label{a5}
\ee
where ${\cal B}=\frac{B}{A(\lambda_R/z-c)}$.

The integral $I(\alpha)$ is a finite function of $v/y$ everywhere, except at
$v/y=0$, where a singularity  develops if $\alpha \geq 1$ and $u \rightarrow 1$. 
In order to see this, let us separate in $I(\alpha)$ the contribution
coming from the neighborhood of the upper limit of integration, by writing $I(\alpha) = I_0 + I_u$,
which corresponds to the partition of the domain of integration 
\be
\int _{1/y}^u  = \int _{1/y}^{u-\epsilon} +\int _{u-\epsilon}^u
\label{a12}
\ee
where $\epsilon$ is a small number. The first contribution $I_0$ is always finite, while 
\begin{eqnarray}
I_u & = & \int _{u-\epsilon}^u dz {z^{\beta + \alpha - (1+a)} \over (1-z +v/y)^\alpha} \\ \nonumber
& \simeq & \frac{u^{\beta + \alpha - (1+a)}}{1-\alpha} [(1-u+\epsilon+v/y)^{1-\alpha} - (1-u+ v/y)^{1-\alpha}]
\label{a13}
\end{eqnarray}
is finite for $u <1$, but diverges for $u=1$ and $v/y \rightarrow \infty$, if $\alpha \geq 1$, yielding
\be
    I(\alpha) \simeq \left \{ \begin{array}{ll}
        < \infty    \qquad   $for$ \qquad \alpha <1   \\ 	
        -\frac{1}{\ln (v/y)}  \qquad $for$ \qquad \alpha =1 \\
        \frac{1}{\alpha -1} (v/y)^{1-\alpha} \qquad $for$ \qquad \alpha >1.
        \end{array}
        \right .
        \label{a131}
\ee
Going through similar steps, one can see that also $I(\alpha+1)$ is always finite for $u <1$, while for $u=1$ and small $v/y$
\be
I(\alpha+1) \simeq \frac{1}{\alpha} (v/y)^{-\alpha}.
\label{a132}
\ee
Therefore, from Eq.~(\ref{a4}) and for $u<1$ we have
\be
\widetilde {F}(y,v) \simeq -(v/y)
\label{a133}
\ee
while for $u=1$
\be
    \widetilde {F}(y,v) \simeq \left \{ \begin{array}{ll}
        -(v/y)^{1-\alpha}   \qquad   $for$ \qquad \alpha <1   \\ 	
        \frac{1}{\ln (v/y)}  \qquad $for$ \qquad \alpha =1 \\
        1-\alpha \qquad $for$ \qquad \alpha >1.
        \end{array}
        \right .
        \label{a134}
\ee
The above results hold in general. In the following we continue separately the discussion of $\omega_{ag}$ and $\pi_{ag}$.

\subsection{$\omega_{ag}$}

With $1 < q \leq y$ and $u=q/y$, the integration domain $[1/y,q/y]$ in Eqs.~(\ref{omega2.1}) and~(\ref{omega2.2}) 
includes $u=1$ only when $y$ takes the smallest possible value $y=q$. Therefore, from Eq.~(\ref{a134}) we obtain
\be
    \widetilde {F}_{\omega}(q,v) \simeq \left \{ \begin{array}{ll}
        -(v/q)^{1-\alpha}   \qquad   $for$ \qquad \alpha <1   \\ 	
        \frac{1}{\ln (v/q)}  \qquad $for$ \qquad \alpha =1 \\
        1-\alpha \qquad $for$ \qquad \alpha >1
        \end{array}
        \right .
        \label{a135}
\ee
while from Eqs.~(\ref{a5}) and~(\ref{a131}) 
\be
     K_{\omega}(q,v) \simeq \left \{ \begin{array}{ll}
        < \infty   \qquad   $for$ \qquad \alpha <1   \\ 	
        -\frac{1}{\ln (v/q)}  \qquad $for$ \qquad \alpha =1 \\
        (v/q)^{\alpha-1} \qquad $for$ \qquad \alpha >1.
        \end{array}
        \right .
        \label{a136}
\ee
Inserting these results into Eq.~(\ref{a00}), for the effective exponent at $y=q$ we get
\be
    \lim_{t_w \to \infty} a_{eff}^\omega (q,t_w) = \left \{ \begin{array}{ll}
        a   \qquad   $for$ \qquad \alpha < 1   \\ 	
        a+1-\alpha = 1/z \qquad $for$ \qquad \alpha \geq 1.
        \end{array}
        \right .
        \label{a137}
\ee
Instead, for $y>q$, since $\widetilde {F}(y,v)$ obeys Eq.~(\ref{a133}) and $K_{\omega}(y,v)$ is finite, we get
\be
\lim_{t_w \to \infty} a_{eff}^\omega (y,t_w) = a
\label{a138}
\ee
independently from the sign of $\alpha - 1$.
However, care must be used when $y \rightarrow \infty$, because the domain of integration in Eq.~(\ref{a2}) shrinks to zero
and $I(\alpha)$ vanishes like
\be
I(\alpha) \sim y^{a-(\beta+\alpha)}
\label{a138bis}
\ee
yielding 
\be
K_{\omega}(y,v) \sim y^{\beta+\alpha - \lambda_R/z}.
\label{a138tris}
\ee
Therefore, from Eq.~(\ref{812}) follows that $K_{\omega}(y,v)$ diverges for large $y$ and, in turn, this gives the
non commutativity of the limits $\lim_{y \to \infty}$ and  $\lim_{t_w \to \infty}$, since keeping $t_w$ fixed
\be
\lim_{y \to \infty} a_{eff}^\omega (y,t_w) = c
\label{a138cicci}
\ee
while with $y > q$ fixed
\be
\lim_{t_w \to \infty} a_{eff}^\omega (y,t_w) = a.
\label{a139}
\ee

\subsection{$\pi_{ag}$}

The integration domain $[1/y,p]$ in Eqs.~(\ref{omega2.1}) and~(\ref{omega2.2}) is given by $[1/y,p]$,
with $ y \geq 1/p$. Therefore, the upper limit of integration $u=1$ is not reached if $p<1$, while it is reached if $p=1$.
Furthermore, as $y$ hits the lowest value $y=1/p$, the integration domain shrinks to zero and to first order in $(y-1/p)$
\be
I(\alpha) = p^{2+ \beta + \alpha - (1+a)}[1-p(1-v)]^{-\alpha}(y-1/p)
\label{a40}
\ee
from which follows
\be
\lim_{y \to 1/p}  \widetilde {F}_{\pi}(y,v)= -\alpha\frac{pv}{1-p+pv}.
\label{a41}
\ee
Notice, also, that from Eq.~(\ref{a5}) we get
\be
K_{\pi}(y,v) =   {{\cal B} \over  I(\alpha)} [p^{\lambda_R/z-c} y^{a-c} - y^{a-\lambda_R/z}]
\label{a42}
\ee
where both the exponents of $y$ in the right hand side are negative. Hence, for fixed $t_w$
\be
\lim_{y \to \infty} K_{\pi}(y,v) = 0
\label{a43}
\ee 
contrary to what happens in the case of $\omega_{ag}$, where the limit for $y \rightarrow \infty$ produces a divergence.
Let us proceed by considering separately $p<1$ and $p=1$.

\vspace{5mm}

{\bf $p<1$}

\vspace{5mm}

\noindent In this case $u < 1$ for all $y$ and $I(\alpha)$ does not develop any singularity. From Eqs.~(\ref{a133}),~(\ref{a41})
and~(\ref{a42}) then follows 
\be 
\lim_{t_w \to \infty} a_{eff}^\pi (y,t_w) = a
\label{a44}
\ee
for all y.

\vspace{5mm}

{\bf $p=1$}

\vspace{5mm}

\noindent From Eq.~(\ref{a41}), at $y=1$ we have
\be
\widetilde {F}_{\pi}(1,v)= -\alpha
\label{a45}
\ee
while for $y>1$ Eq.~(\ref{a134}) applies. Switching to the behavior of $K_{\pi}(y,v)$, notice that inserting Eq.~(\ref{a131}) into Eq.~(\ref{a42}) we obtain
\be
     K_{\pi}(y,v) \simeq \left \{ \begin{array}{ll}
         y^{a-c} - y^{a-\lambda_R/z}  \qquad   $for$ \qquad \alpha <1   \\ 	
        -\frac{ y^{a-c} - y^{a-\lambda_R/z}}{\ln (v/q)}  \qquad $for$ \qquad \alpha =1 \\
         ( y^{a-c} - y^{a-\lambda_R/z})(v/q)^{\alpha-1} \qquad $for$ \qquad \alpha >1.
        \end{array}
        \right .
        \label{a46}
\ee
Using these results, for fixed $y$ we get
\be
    \lim_{t_w \to \infty} a_{eff}^\pi (y,t_w) = \left \{ \begin{array}{ll}
        a-\alpha = 1/z -1   \qquad   $for$ \qquad y=1   \\ 
	a \qquad   $for$ \qquad y>1 \qquad $and$ \qquad  \alpha < 1 \\
        a+1-\alpha = 1/z \qquad $for$ \qquad   y>1 \qquad $and$ \qquad   \alpha \geq 1
        \end{array}
        \right .
        \label{a47}
\ee
while for fixed $t_w$
\be
    \lim_{y \to \infty} a_{eff}^\pi (y,t_w) = \left \{ \begin{array}{ll}
        a \qquad   $for$ \qquad  \alpha < 1 \\
        a+1-\alpha = 1/z \qquad $for$ \qquad   \alpha \geq 1.
        \end{array}
        \right .
        \label{a48}
\ee
Finally, let us recall that from Eqs.~(\ref{aphen}) and~(\ref{alp}) we have
\be
\alpha -1 = \frac{1}{z} \frac{d -d_U}{d_U -d_L}
\label{a49}
\ee
which shows how the sign of $\alpha -1$ changes with the dimensionality.

\vspace{2cm}

{\bf Acknowledgements}
 
This work has been partially supported by MURST through PRIN-2004.

\newpage

\newpage

\begin{thebibliography}{99}
\bibitem{cop}
This is no longer correct in the case of a conserved order parameter~\cite{yeung}.
\bibitem{nota*}
$d_U=2$ in the approximate calculations~\cite{epj,Berthier,Mazenko04} for systems with scalar order parameter show that
there are quantitative features of $R_{ag}(t,s)$ beyond the reach of the best analytical methods presently available.
\bibitem{nota4}
In the first set we have omitted $s=102,103,104$ since these values are too close to $s=101$.
\end{thebibliography}
\end{document}